\documentclass[aps,prd,preprintnumbers,showpacs,showkeys,nofootinbib,
superscriptaddress,fleqn,floatfix,tightenlines,10pt]{revtex4-1}
\usepackage{amsmath,amsfonts,amssymb,amscd,amsxtra,amsthm}
\usepackage{graphicx}  
\usepackage{epstopdf}
\usepackage{dcolumn}  
\usepackage{bm}          
\usepackage{slashed}
\usepackage{cancel}
\usepackage{float}
\usepackage{mathtools}
\usepackage{amsbsy}
\usepackage{amstext}

\usepackage[utf8]{inputenc}
\usepackage{booktabs}
\usepackage[normalem]{ulem} 
\usepackage[dvipsnames]{xcolor} 
\usepackage{tabularx}
\usepackage{enumitem}
\usepackage{array}
\usepackage{slashed} 
\usepackage{tikz}
\usepackage{float}
\usepackage{multirow}
\renewcommand\sout{\bgroup \color{red} \ULdepth=-.5ex \ULset}

\makeatletter

\begin{document}
\preprint{INHA-NTG-13/2019}

\title{The $\sigma$ and $\rho$ coupling constants for the charmed and
  beauty mesons} 

\author{Hee-Jin Kim}
\email{heejin.kim@inha.edu}
\affiliation{Department of Physics, Inha University, Incheon 22212,
  Korea}

\author{Hyun-Chul Kim}
\email{hchkim@inha.ac.kr}
\affiliation{Department of Physics, Inha University, Incheon 22212,
  Korea}
\affiliation{School of Physics, Korea Institute for Advanced Study
  (KIAS), Seoul 02455, Republic of Korea}

\date{June, 2020}

\begin{abstract}
We investigate the $\sigma$ and $\rho$ coupling constants for the
$DD$ and $D^*D^*$ interactions, based on correlated $2\pi$ exchange
in the $DD$ and $D^*D^*$ interactions. Starting from the
$D\bar{D}\to \pi\pi$ and $D^*\bar{D}^*\to \pi\pi$ amplitudes
derived in the pseudophysical region ($4m_\pi^2\le t \le 52m_\pi^2$)
with the $S$- and $P$-wave $2\pi$ correlations considered, we obtain
the spectral functions for the $DD\to DD$ and $D^*D^*\to D^*D^*$
amplitude with correlated $S$- and $P$-wave $2\pi$
exchanges. Using the pole approximation, we estimate the $DD\sigma$,
$DD\rho$, $D^*D^*\sigma$, and $D^*D^*\rho$ coupling constants. 
We extended phenomenologically the present results to the region in
$t\le 0$ and compare them with those from lattice QCD.
The results are also compared with those of other models. We also
present the results of the $BB\sigma$, $BB\rho$, $B^*B^*\sigma$, and
$B^*B^*\rho$ coupling constants. We observe that it is unlikely that
the $\sigma$ and $\rho$ coupling constants for the $B$ and $B^*$
mesons are the same as those for the $D$ and $D^*$ mesons. On the 
contrary, they are quite larger than those for the charmed mesons.
\end{abstract}


\maketitle

\section{Introduction}
Understanding exotic heavy mesons has been one of the most important
issues in hadronic physics (see, for
example, recent reviews~\cite{Swanson:2006st, Chen:2016qju,
  Lebed:2016hpi, Olsen:2017bmm, Guo:2017jvc}). In 2003, a
charmonium-like state $X(3872)$ was newly found by the Belle
Collaboration~\cite{Choi:2003ue} and was subsequently confirmed by 
other experiments~\cite{Acosta:2003zx,Abazov:2004kp, Aaij:2011sn,
  Chiochia:2012cg}. The mass of the $X(3872)$ turns out to be
approximately few tens of MeV lower than the $P$-wave charmonium
$\chi_{c1}(2P)$ predicted by the heavy-quark potential models
\cite{Eichten:1980mw, Godfrey:1985xj, Gupta:1986xt, Fulcher:1991dm,
  Zeng:1994vj, Ebert:2002pp}. The Belle
Collaboration~\cite{Choi:2003ue} also measured the upper limit on the
ratio of the partial decay widths $\Gamma(X(3872)\to
\gamma\chi_{c1})/\Gamma(X(3872)\to \pi^+\pi^- J/\psi)<0.89$, which was
very different from the predictions of Ref.~\cite{Eichten:2002qv} in
which the decays of $D$-wave missing charmoniums were considered.
Though there was an argument that the 
$X(3872)$ is still a $1{}^3D_2$ charmonium
state~\cite{Pakvasa:2003ea}, these discrepancies led to various
theoretical interpretations on the $X(3872)$. It can be regarded as a
tetraquark state~\cite{Maiani:2004vq,Hogaasen:2005jv,Ebert:2005nc} or
as a hybrid exotic state~\cite{Li:2004sta}. It is also plausible to
consider it as a molecular state of $D$ and $\bar{D}^{*}$
mesons~\cite{Tornqvist:1993ng, Tornqvist:2003na, Close:2003sg,
  Voloshin:2003nt, Wong:2003xk}, since its mass is very close to the
sum of the masses of the $D$ and $D^*$ mesons
($M_{X(3872)}-M_{D^{*0}}-M_{D^0}=0.01 \pm 0.18$ MeV), which resembles
the deuteron consisting of the proton and the neutron. The quantum
number of the $X(3872)$ is now established as
an isosinglet state with $J^{PC}=1^{++}$~\cite{Aaij:2015eva}.

In addition to the $X(3872)$ meson, a number of new heavy mesons has
been experimentally observed over the last decade (see, for example,
glossary of exotic states summarized in Appendix of
Ref.~\cite{Lebed:2016hpi}). Many of them can be regarded as molecular
states. While the pion exchange is a main ingredient for the
description of those exotic mesons as molecular states, the $\sigma$
meson exchange may come into play, since it provides a strong
attraction in medium range of the interaction so as to make two heavy
mesons such as $D$ ($D^*$) and $\bar{D}^*$ ($D^*$) bound. However, the
coupling constants for the $DD\sigma$ and $D^*D^*\sigma$ vertices are
not well known both theoretically and experimentally, so that
these couplings have been estimated by using either the nonlinear
sigma model or quark models~\cite{Liu:2008tn, Lee:2009hy, Liu:2010xh,
  Liu:2019stu}. 
Moreover, the $D^*D^*\sigma$ coupling constant was taken to be the
same as the $DD\sigma$ one in many theoretical works with the
heavy-quark spin symmetry assumed. On the other hand, the $\sigma$
exchange in the $NN$ interaction is known to be a parametrization of
the correlated $2\pi$ exchange, based on the pole
approximation~\cite{Brown, Machleidt:1987hj, Kim:1994ce}, which
approximates the broad mass distribution of the $\sigma$ meson to a
sharp mass. This $\sigma$-exchange contribution has been an essential
part of providing the strong attraction in the intermediate range of
the $NN$ potential. In fact, the $S$-wave correlated $2\pi$ exchange
has been employed in predicting the $DD$ and $BB$ bound
states~\cite{Kim:1995bm} already. Since one $\pi$ exchange is not
allowed between the pseudoscalar heavy mesons, $\sigma$ exchange plays
a crucial role in examining the bound states of the $DD$ and $BB$
system. Similarly, $\rho$-meson exchange can be also regarded as a
parametrization of the correlated $2\pi$ exchange in the $P$-wave
(vector-isovector) channel~\cite{Machleidt:1987hj,Kim:1994ce}. 

In the present work, we derive the five different coupling constants:
$g_{DD\sigma}$, $g_{DD\rho}$, $g_{D^*D^*\sigma}$, $g_{D^*D^*\rho}$,
and $f_{D^*D^*\rho}$, based on the pseudophysical $D\bar{D}\to \pi\pi$
and $D^*\bar{D}^* \to \pi\pi$ amplitudes. The $NN\sigma$ coupling
constant can be determined by using the pseudophysical $N\bar{N}\to
\pi\pi$ amplitudes in the $NN$ interaction. In
Ref.~\cite{Durso:1980vn}, the $N\bar{N}\to\pi\pi$ amplitudes in the
pseudophysical region with $\pi\pi$ rescattering was constructed, from
which the $NN\sigma$ coupling constant with broad width  can be
extracted. In this work, we closely follow the theoretical technique
developed in Ref.~\cite{Durso:1980vn}. In fact, the same method was
adopted in the full Bonn potential for the $NN$ interaction, $\sigma'$
exchange being replaced by the \textit{correlated} $2\pi$ exchange
developed in this way~\cite{Kim:1994ce}. Later, this approach was also
employed in the J{\"u}lich-Bonn potential for the hyperon-nucleon
interaction~\cite{Reuber:1995vc}.  Thus, we will take this
well-established method to determine the $\sigma$ coupling constants
for the $D$ and $D^*$. We will also apply this
to the $\sigma$ coupling constants for the $B$ and $B^*$ mesons.

\begin{figure}[htp]
\centering
\includegraphics[scale=0.55]{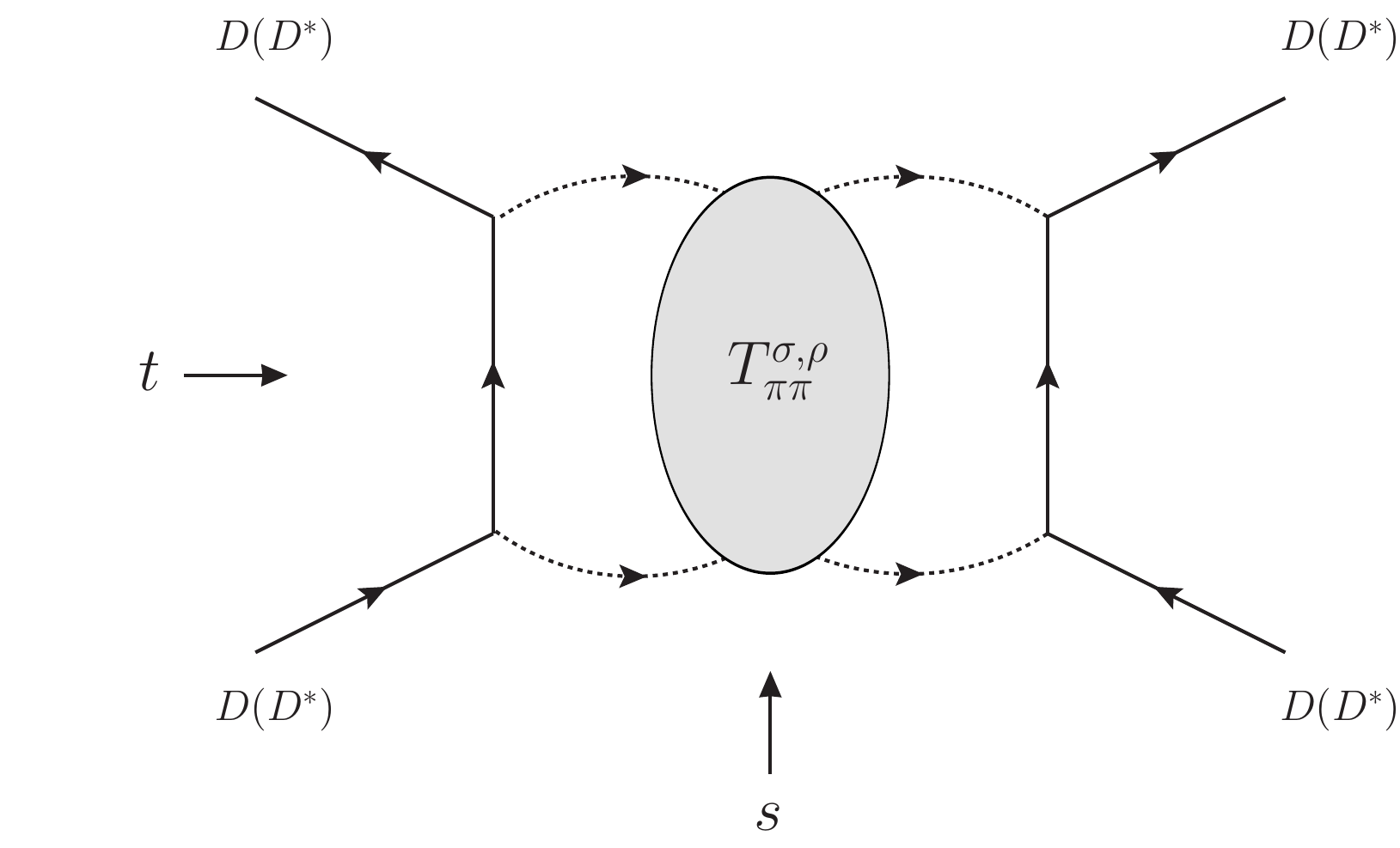}
  \caption{Schematic diagram for the correlated $2\pi$ exchange.}
  \label{fig:1}
\end{figure}

The schematic diagram for the correlated $2\pi$
exchange in the $DD$ and $D^*D^*$ interactions is drawn in
Fig.~\ref{fig:1}.
To determine these coupling constants, we first formulate the
off-shell $D\bar{D}\to \pi\pi$ and $D^*\bar{D}^* \to \pi\pi$
amplitudes in the pseudophysical region ($4m_\pi^2 \le t \le 52
m_\pi^2$), where $t$ is the total energy squared in
the center of momentum (CM) system, using the effective
Lagrangians. We want to mention that there is one caveat. The kaon and
antikaon ($K\bar{K}$) channel is open at around $50\,m_\pi^2$. Thus,
it is natural to include the $K\bar{K}$ channel in a coupled-channel 
formalism. However, since there is no information on the relevant
coupling constants, it is inevitable to introduce additional
uncertainties in the present calculation by including the $K\bar{K}$
channel. Thus, we will consider only the $\pi\pi$ channel.  
Then, we combine the $D\bar{D}\to \pi\pi$ amplitudes with
the off-shell   $\pi\pi$ amplitudes evaluated within the J{\"u}lich
$\pi\pi$ model~\cite{Lohse:1990ew,Janssen:1994wn} but
modified in a covariant way. The model described very well the phase
shifts of $\pi\pi$ scattering in both the scalar-isoscalar and
vector-isovector channels. The $\pi\pi$ amplitudes with
meson-exchange picture is the most consistent and convenient one for
the present approach, because we will construct the off-shell Born
amplitudes for the $D\bar{D}\to\pi\pi$ and $D^*\bar{D}^*\to\pi\pi$
based on the effective Lagrangians. We want to mention that 
the method we adopt in this work is basically the same as 
shown in Ref.~\cite{Durso:1980vn, Kim:1994ce}. The two-body unitarity
allows one to construct the spectral functions for the $D\bar{D}$ and
$D^*\bar{D}^*$ amplitudes. Having derived these spectral functions,
one can directly compute the coupling constants listed above by using
the dispersion relation. For completeness, we also present the results
of the $\sigma$ and $\rho$ coupling constants for the $B$ and $B^*$
mesons. 

The present work is organized as follows: In Section II, we show how
to derive the spectral functions from which the coupling constants can
be determined. In Section III, we discuss the present results in
comparison with those of other works. The last Section is devoted to
summary and conclusion.

\section{General formalism}
\subsection{$D\bar{D}\to \pi\pi$ and $D^*\bar{D}^*\to \pi\pi$ amplitudes}
We start with the effective Lagrangian from HQET~\cite{Wise:1992hn,
  Wise:1993wa, Yan:1992gz, Wise:1993wa, Burdman:1992gh}
\begin{align}
\mathcal{L} \;=\; i g\mathrm{tr}\left[H_b \rlap{/}{A}_{ba} \gamma_5
  \bar{H}_a\right],
\end{align}
where the heavy meson field $H_b$ is given as
\begin{align}
H_b \;=\; \frac{1+\rlap{/}{v}}{2} \left[ P_b^{*\mu} \gamma_\mu - P_b
  \gamma_5\right] ,
\end{align}
and the axial-vector field $A_{ba}^\mu$ is expressed as
\begin{align}
A_{ba}^\mu \;=\; \frac12 (\xi^\dagger \partial^\mu \xi -
\xi \partial^\mu \xi^\dagger) \;=\; \frac{i}{f_\pi} \partial^\mu
\mathcal{M}_{ba} + \cdots
\end{align}
with the pseudoscalar meson field $\mathcal{M}$
\begin{align}
\mathcal{M} \;=\; \left(
  \begin{array}{ccc}
\frac{\pi^0}{\sqrt{2}}+\frac{\eta}{\sqrt{6}} & \pi^+ & K^+ \\
\pi^- &  -\frac{\pi^0}{\sqrt{2}}+\frac{\eta}{\sqrt{6}}  & K^0 \\
K^- & \bar{K}^0 & -\frac{2\eta}{\sqrt{6}}
  \end{array}
\right).
\end{align}
The pseudo-Nambu-Goldstone (pNG) field or the coset field
$\xi(x)$~\cite{Callan:1969sn}, which realizes the emergence of the
pNG field due to the spontaneous breakdown of chiral symmetry, is
defined as 
\begin{align}
\xi(x) = \exp[i \mathcal{M}(x)/f_\pi]
\end{align}
with the pion decay constant $f_\pi=132$ MeV normalized.  Since we
need only the single pNG field, we keep the
expansion to linear order with respect to $\mathcal{M}$.
The Dirac conjugate of the heavy meson field $\bar{H}_a$ is
written as $\bar{H}_a \;=\; \gamma_0 H^\dagger \gamma_0$.
The effective Lagrangians for $PP^*M$ and $P^ *P^* M$ couplings are
then 
\begin{align}
\mathcal{L}_{PP^*M} &= -\frac{2g}{f_\pi} P^{*\mu} \partial_\mu
\mathcal{M}_{ba} P^\dagger + \mbox{h.c.},\cr
\mathcal{L}_{P^*P^*M} &=  \frac{2gi}{f_\pi}
                        P_b^{*\beta} \partial^\mu
\mathcal{M}_{ba} P_a^{*\alpha\dagger}\varepsilon_{\alpha\beta\mu\nu} v^\nu,
\label{eq:Lagrangian}
\end{align}
where $f_\pi$ denotes the pion decay constant of which the value is
taken from the experimental one, $f_\pi=132$ MeV. Note that the parity
conservation does not allow the $DD\pi$ vertex. 

\begin{figure}[htp]
\centering
\includegraphics[scale=0.75]{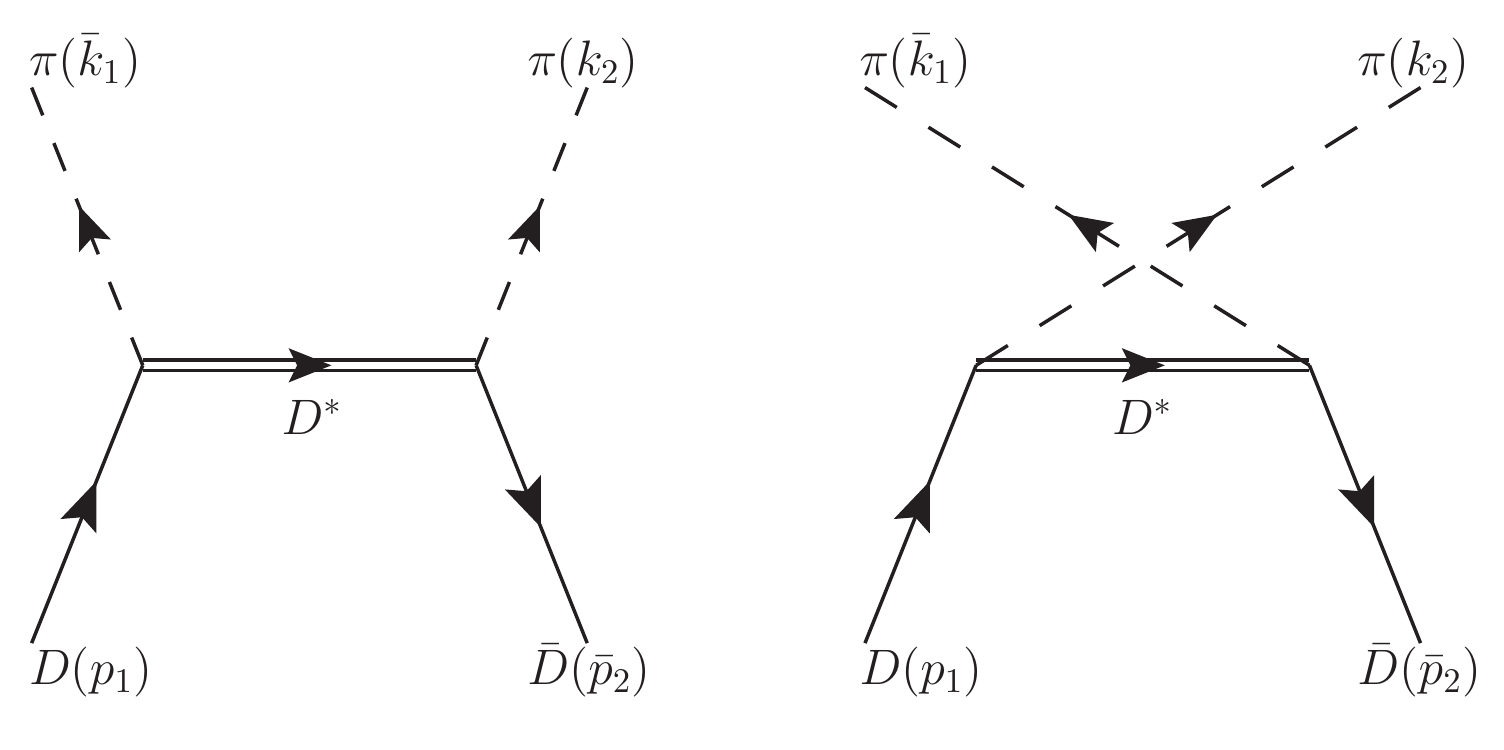}
  \caption{Feynman diagrams for $ D\bar{D} \to \pi \pi$}
  \label{fig:2}
\end{figure}
\begin{figure}[htp]
\centering
\includegraphics[scale=0.75]{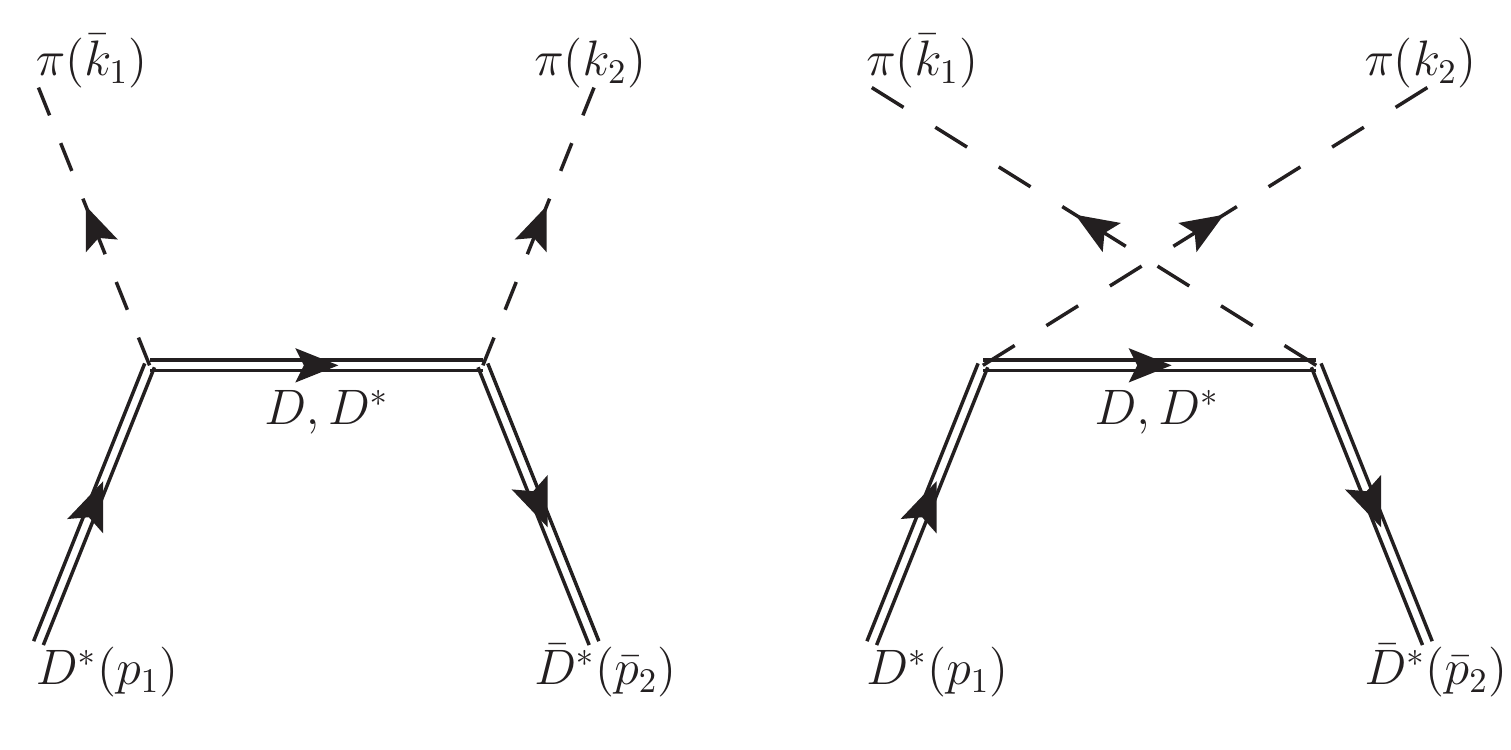}
  \caption{Feynman diagrams for $ D^*\bar{D}^* \to \pi \pi$}
  \label{fig:3}
\end{figure}
Using Eqs.~\eqref{eq:Lagrangian}, we can compute the off-shell Born
amplitudes for the $D\bar{D}\to \pi\pi$ and $D^*\bar{D}^* \to \pi\pi$
processes. The corresponding Feynman diagrams are given in
Fig.~\ref{fig:2}. As for the $D\bar{D}\to \pi\pi$
process, we need to consider only $D^*$-meson exchange, whereas we
have to consider both $D$- and
$D^*$-meson exchanges for the $D^*\bar{D}^*\to \pi\pi$ process.
As depicted in Figs.~\ref{fig:2} and~\ref{fig:3},
$p_1$ and $\bar{p}_2$ stand for four-momenta for the initial $D$ ($D^*$) and
$\bar{D}$ ($\bar{D}^*$) mesons, and $\bar{k}_1$ and $k_2$ denote those
for the final pions, respectively. In the CM frame,
they are expressed as
\begin{eqnarray}
p_1 &=& (E_p,\,\bm p),\;\;\; \bar{p}_2 \;=\; (E_p, -\bm p),\cr
\bar{k}_1 &=& (\omega_k,\,\bm k),\;\;\; k_2 \;=\; (\omega_k,-\bm k),
\end{eqnarray}
where
\begin{align}
E_p \;=\; \sqrt{M_{D(D^*)}^2 + \bm p^2},\;\;\;
\omega_k \;=\; \sqrt{m_\pi^2 + \bm k^2}.
\end{align}
The conservation of the total momentum is given as
\begin{align}
p_1 + \bar{p}_2 \;=\; \bar{k}_1 + k_2.
\end{align}
The Mandelstam variables in the $t$ channel are defined as
\begin{align}
s &= (p_1-\bar{k}_1)^2 \;=\; (\bar{p}_2-k_2)^2\;=\;p_1^2 + \bar{k}_1^2
    -2 p_1\cdot     \bar{k}_1, \cr
t &= (p_1+\bar{p}_2)^2 \;=\; (\bar{k}_1+k_2)^2 = p_1^2 + \bar{p}_2^2 +
    2 p_1\cdot \bar{p}_2 \;=\; \bar{k}_1^2 + k_2^2 + 2\bar{k}_1\cdot
    k_2,\cr
u &= (p_1-k_2)^2 \;=\; (\bar{p}_2-\bar{k}_1)^2 = \bar{p}_2^2 +
    \bar{k}_1^2 - 2 \bar{p}_2\cdot \bar{k}_1,
\end{align}
where $t$ is just the square of the total energy while $s$ represents
the squre of the momentum transfer.
Since we are interested in the pseudophysical region and want to
combine these Born amplitudes with the $\pi\pi$ rescattering
amplitude, we have to consider the off-mass shell pion states. Thus,
we need to take into account the virtual momenta of the two pions,
i.e.,
\begin{align}
\bar{k}_1^2 = \frac{t}{4} - \bm k^2 ,\;\;\; k_2^2 = \frac{t}{4} - \bm
  k^2, \;\;\; \bar{k}_1\cdot k_2 \;=\; \frac{t}{4} + \bm k^2.
\end{align}
Since we will use the $D\bar{D}\to \pi\pi$ ($D^*\bar{D}^*\to \pi\pi$)
amplitudes to derive the
spectral functions of $D\bar{D}\to D\bar{D}$ ($D^*\bar{D}^*\to
D^*\bar{D}^*$) with $2\pi$ unitarity, we need to consider the off-mass
shell momenta for pions. The sum of all the Mandelstam variables are
expressed in terms of the masses of $D$ and $\pi$
\begin{align}
s+t+u  \;=\; 2(M_{D(D^*)}^2+ m_\pi^2).
\end{align}
with the conservation of the total momentum. Let $\theta$ be the angle
between $\bm p$ and $\bm k$. Then,
\begin{align}
s &= M_{D(D^*)}^2 + \frac t{4} - \bm k^2 - 2 E_p \omega_k
      + 2 |\bm p||\bm k|\cos\theta ,\cr
t &= 4\omega_k^2\,\ge 4 m_\pi^2 ,\cr
u &= 2M_{D(D^*)}^2 + 2m_\pi^2 - s - t.
\end{align}

The invariant amplitudes for the $D\bar{D}\to\pi\pi$ process are given
as
\begin{align} \label{eq:d}
\mathcal{M}_{\alpha\beta}^s(D\bar{D}\to \pi\pi) &=
g_{DD^*\pi}^2 \frac{\bar{k}_1 \cdot k_2 -
\frac{(\bar{k}_1 \cdot q)(k_2 \cdot q)}{M_{D^*}^2}}{s-M_{D^*}^2}
\tau_\alpha \tau_\beta,\cr
\mathcal{M}_{\alpha\beta}^u(D\bar{D}\to \pi\pi) &=
-g_{DD^* \pi}^2 \frac{\bar{k}_1 \cdot k_2 -
\frac{(\bar{k}_1 \cdot q)(k_2 \cdot q)}{M_{D^*}^2}}{u-M_{D^*}^2}
\tau_\beta \tau_\alpha.
\end{align}
Similarly, those for the $D^*\bar{D}^*\to\pi\pi$ process are obtained as
\begin{align} \label{eq:dstar}
\mathcal{M}_{\alpha \beta}^{s}(D^*\bar{D}^*\to \pi\pi) & = \left[
g_{DD^* \pi}^2 \frac{(\epsilon^{(\lambda)}
(p_1)\cdot \bar{k}_1)( \epsilon^{(\lambda')}(\bar{p}_2)\cdot
k_2)}{s - M_D^2} \right. \cr
&\left.+\, 4g_{D^*D^* \pi}^2 \varepsilon_{\mu_1 \nu_1 \rho_1 \sigma_1}
\varepsilon_{\mu_2 \nu_2 \rho_2 \sigma_2} g^{\mu_1 \mu_2}
\frac{\epsilon^{(\lambda) \nu_1} (p_1) \epsilon^{(\lambda') \nu_2} (p_2)
p_1^{\sigma_1} \bar{p}_2^{\sigma_2}
\bar{k}_1^{\rho_1} k_2^{\rho_2}}{s-M_{D^*}^2}
\right] \tau_\alpha \tau_\beta
,\cr
\mathcal{M}_{\alpha \beta}^{u}(D^*\bar{D}^*\to \pi\pi) & =
 \left[g_{DD^* \pi}^2 \frac{(\epsilon^{(\lambda)} (p_1)\cdot k_2)(
\epsilon^{(\lambda')}(\bar{p}_2)\cdot  \bar{k}_1)}{u - M_D^2} \right. \cr
&\left.+\, 4g_{D^*D^* \pi}^2 \varepsilon_{\mu_1 \nu_1 \rho_1 \sigma_1}
\varepsilon_{\mu_2 \nu_2 \rho_2 \sigma_2} g^{\mu_1 \mu_2}
\frac{\epsilon^{(\lambda) \nu_1} (p_1) \epsilon^{(\lambda') \nu_2} (p_2)
p_1^{\sigma_1} \bar{p}_2^{\sigma_2}
k_2^{\rho_1} \bar{k}_1^{\rho_2}}{u-M_{D^*}^2}
\right]  \tau_\beta \tau_\alpha.
\end{align}
The total amplitude can be generically expressed in terms of
the iso-symmetric amplitude $\mathcal{M}^{(+)}$ and the
iso-antisymmetric amplitude $\mathcal{M}^{(-)}$
\begin{align}
\mathcal{M}_{\alpha\beta} \;=\; \mathcal{M}^{(+)} \delta_{\alpha\beta}
+ \mathcal{M}^{(-)} \frac12[\tau_\alpha,\,\tau_\beta] ,
\end{align}
where
\begin{align}
\mathcal{M}^{(+)} =  \mathcal{M}^{s} + \mathcal{M}^u,\;\;\;
\mathcal{M}^{(-)} =  \mathcal{M}^{s} - \mathcal{M}^u.
\end{align}
Note that the isospin amplitudes for $J=0$ and $J=1$ are respectively
related to $\mathcal{M}^{(+)}$ and $\mathcal{M}^{(-)}$ as in the case
of the $N\bar{N}\to\pi\pi$ amplitudes~\cite{Frazer:1960zza,
  Kim:1994ce}
\begin{align}
\mathcal{M}^{(+)} = -\frac1{\sqrt{6}} \mathcal{M}_{T=0},\;\;\;
\mathcal{M}^{(+)} = -\frac1{2} \mathcal{M}_{T=1}.
\end{align}

In order to compute the amplitudes given in Eqs.~\eqref{eq:d}
and~\eqref{eq:dstar}, we have to introduce the form factors at each
vertex. We employ a Gaussian-type form factor defined as
\begin{align}
F(s) =
  \exp\left[\frac{s-M_{\mathrm{ex}}^2}{\Lambda_{\mathrm{ex}}^2}
  \right],
\end{align}
where $M_{\mathrm{ex}}$ represents either the mass of $D$ meson or
that of the $D^*$ meson. $\Lambda_{\mathrm{ex}}$ denotes the cutoff
mass corresponding to the exchange particle. In
Ref.~\cite{Kim:1994ce}, the value of the cutoff mass $\Lambda_N$
($\Lambda_{\Delta}$) was taken to be
around 1.5 GeV (1.7 GeV) when the spectral functions for the
$NN\sigma$ and $NN\rho$ coupling constants were investigated. However,
the masses of the $D$ and $D^*$ mesons are approximately two times
larger than the nucleon or the $\Delta$ isobar, the values of the
cutoff mass for the $DD\pi$ and $D^*D^*\pi$ vertices should be taken
to be larger than those for the $NN\pi$ and $N\Delta \pi$
ones. Otherwise, the amplitudes depend too sensitively on the
cutoff masses, because their numerical values are too close to the
masses of the exchanged $D$ or $D^*$ mesons. Moreover, 
a recent work on the electromagnetic form factors of the heavy
baryons~\cite{Kim:2018nqf} has shown that the heavy baryon is a much
more compact object in comparison with the proton. It indicates that
the cutoff mass of the parametrized heavy-baryon electric form factor
should be larger than that of the proton at least by about a factor
1.6, because the cutoff mass is implicitly related to the
corresponding particle. Thus, we will use the values of 
the cutoff masses $\Lambda_D = 2.5$ GeV and $\Lambda_{D^*} = 2.8$
GeV in the present work. Moreover, if one uses smaller values of the
cutoff masses, one can not get stable results for the form factors
corresponding to the coupling constants in the physical $t$ region
($t\le 0$). At a certain value of $-t$, the $\sigma$ coupling constant
even vanishes, which leads to unphysical results. 

As far as an explicit form of the form
factors is concerned, one could utilize the well-known monopole- or
dipole-type form factor. However, we find that such types of the form
factors do not suppress enough the Born amplitudes as the total energy
increases. For example, the Born amplitudes for the $D^*\bar{D}^*\to
\pi\pi$ process given in Eq.~\eqref{eq:dstar} has a strong dependence
on $t$. In particular, the second terms in the $s$- and $u$-channel
amplitudes contain the four momenta in the numerator, which make the
amplitudes too large as $t$ increases. To tame this behavior,
we employ the Gaussian-type form factor at each vertex.

\subsection{Rescattering equation and spectral functions}
Once the off-shell Born amplitudes have been evaluated, the next step
is to compute the rescattering equation that combines
the off-shell $D\bar{D}\to \pi\pi$ and $D^*\bar{D}^*\to \pi\pi$
amplitudes with the off-shell $\pi\pi$ amplitudes.
\begin{figure}[htp]
\centering
\includegraphics[scale=0.55]{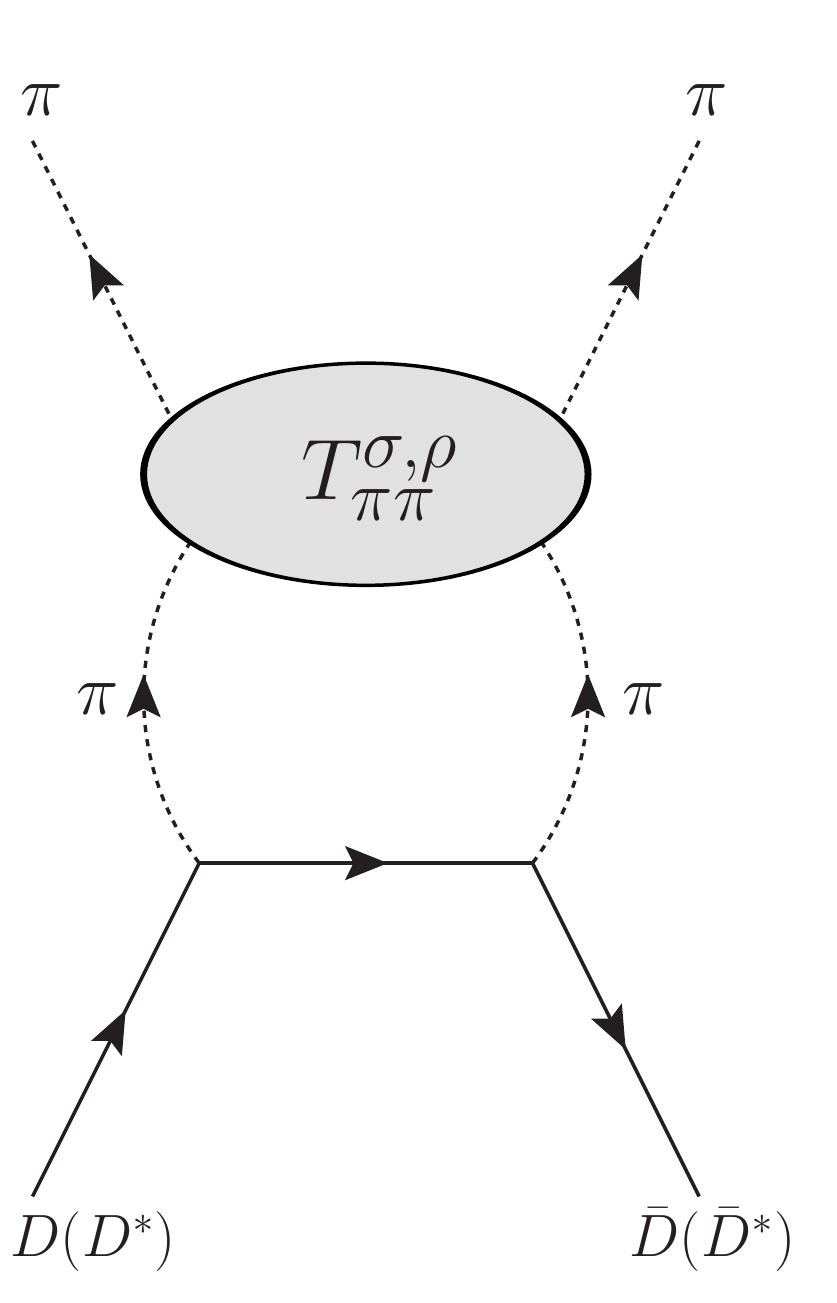}
  \caption{Rescattering equation}
  \label{fig:4}
\end{figure}
As shown in Fig.~\ref{fig:4}, we need to incorporate the $\pi\pi$
interaction in the course of the $D\bar{D}\to\pi\pi$ and
$D^*\bar{D}^*\to\pi\pi$ processes. This can be achieved by
considering the Blankenbecler-Sugar (BbS)
equation~\cite{Blankenbecler:1965gx, Durso:1980vn}, which was derived
by the three-dimensional reduction of the Bethe-Salpeter equation:
\begin{align}
  \label{eq:BS}
\mathcal{M}_{D\bar{D}(D^*\bar{D}^*)\to\pi\pi} (p,p';t) \;=\;
\mathcal{M}_{D\bar{D}(D^*\bar{D}^*)\to\pi\pi}^{\mathrm{Born}}
  (p,p';t)  + \frac1{(2\pi)^3}\int d^3 q \frac1{ \omega_q}
  \frac{\mathcal{M}_{D\bar{D}(D^*\bar{D}^*)\to\pi\pi} (p,q;t)
  \tau(q,p';t)}{t- 4\omega_q^2 + i\varepsilon}.
\end{align}
Since we consider only the scalar-isoscalar ($J=0,\, T=0$) and
vector-isovector channels ($J=1,\,T=1$), which will provide the
$DD(D^*D^*)\sigma$- and $DD(D^*D^*)\rho$-meson coupling constants
respectively,  we can make a partial-wave expansion of the amplitudes
so that we get the partial-wave rescattering equation for the
partial-wave amplitudes with $J$ and $T$ given: 
\begin{align}
\mathcal{M}_{JT}^{D\bar{D}(D^*\bar{D}^*)\to\pi\pi} (t) \;=\;
\mathcal{M}_{JT}^{D\bar{D}(D^*\bar{D}^*)\to\pi\pi,\mathrm{Born}}
  (t) + \frac1{2\pi^2} \int dq q^2
  \frac{\mathcal{M}_{JT}^{D\bar{D}(D^*\bar{D}^*)\to\pi\pi} (p,q;t)
  \tau_{JT}^{\pi\pi} (q,p';t)}{2\omega_q (t-
  4\omega_q^2 +  i\varepsilon)},
\end{align}
where $\tau_{JT}^{\pi\pi}$ denote the off-shell $\pi\pi$ amplitudes,
which were taken from the J{\"u}lich $\pi\pi$ scattering
model~\cite{Lohse:1990ew,Janssen:1994wn}\footnote{One could argue
that a modern $\pi\pi$ amplitude would serve better for the present
work. While the modern $\pi\pi$ amplitudes developed in chiral
perturbation theory with the Roy-like equations are certainly
theoretically more rigorous, the present $\pi\pi$ amplitudes taken
from the meson-exchange picture are consistent and well fitted in 
the present approach and furthermore provide the full
off-mass-shellness that is essential in solving the rescattering
equations.}.  

Next, we consider the unitarity of the  $S$-matrix, i.e.,
\begin{align}
SS^\dagger \;=\; S^\dagger S \;=\; 1.
\end{align}
Since the $S$-matrix for two-body processes is expressed in terms of
the $\mathcal{M}$-matrix or the Feynman invariant amplitude
\begin{align}
S_{fi} \;=\; \delta_{fi} + i(2\pi)^4\delta(p_3+p_4-p_1-p_2)
\mathcal{M}_{fi},
\end{align}
we can get the unitary relation
\begin{align}
S_{fn}S_{ni}^\dagger &= \delta_{fi} + i (2\pi)^4 \delta^{(4)}(P_f-P_i)
                       (\mathcal{M}_{fi}
-\mathcal{M}_{fi}^\dagger ) + (2\pi)^{8} \delta^{(4)}(P_f-P_i) \sum_n
\mathcal{M}_{fn} \mathcal{M}_{ni}^\dagger \delta^{(4)}(P_f-P_n)
= \delta_{fi},
\end{align}
which leads to the following relation
\begin{align}
2\,\mathrm{Im} \mathcal{M}_{fi} \;=\; (2\pi)^4 \sum_n \delta^{(4)}(P_f
- P_n) \mathcal{M}_{fn} \mathcal{M}_{ni}^\dagger ,
\end{align}
where the summation runs over the $2\pi$ states, which is often called
two-body unitarity. More explicitly, we can write it as
\begin{align}
 2\,\mathrm{Im}   \mathcal{M}_{D\bar{D}(D^*\bar{D}^*)} \;=\; (2\pi)^4
  \sum_n   \delta^{(4)}(P_f - P_n)
  \mathcal{M}_{D\bar{D}(D^*\bar{D}^*)\to\pi\pi}
  \mathcal{M}_{\pi\pi\to D\bar{D}(D^*\bar{D}^*)}^\dagger  .
\end{align}

Since we consider the two-body intermediate states, the unitarity
relation can be explicitly written as
\begin{align}
2\,\mathrm{Im}   \mathcal{M}_{D\bar{D}(D^*\bar{D}^*)} = (2\pi)^4
  \frac1{2!} \int  \frac{d^3 q_1}{(2\pi)^3 2 \omega_1} \frac{d^3
  q_2}{(2\pi)^3 2\omega_2}
\delta^{(4)}(p_1+p_2-q_1-q_2)
  \mathcal{M}_{D\bar{D}(D^*\bar{D}^*)\to\pi\pi}
\mathcal{M}_{\pi\pi\to   D\bar{D}(D^*\bar{D}^*)}^\dagger .
\end{align}
Note that $1/2!$ is introduced because of the Bose symmetry. Having
carried out the integrals, we obtain 
\begin{align}
\,\mathrm{Im}   \mathcal{M}_{D\bar{D}(D^*\bar{D}^*)} = \frac1{128\pi^2}
  \sqrt{\frac{t-4m_\pi^2}{t}} \int d\Omega\,
|\mathcal{M}_{D\bar{D}(D^*\bar{D}^*)\to\pi\pi}|^2 .
\end{align}
Taking into account the isospin factors, we find the following
relation
\begin{align}
\mathcal{M}_{D\bar{D}(D^*\bar{D}^*)} =
  3\mathcal{M}_{D\bar{D}(D^*\bar{D}^*)}^{(+)} + 2
  \mathcal{M}_{D\bar{D}(D^*\bar{D}^*)}^{(-)} \bm{\tau}_1\cdot \bm{\tau}_2.
\end{align}

In fact, we need to make a partial-wave expansion in the unitarity
relation, since we want to extract the scalar-isoscalar ($\sigma$) and
vector-isovector ($\rho$) channels from the $D\bar{D}\to
D\bar{D}$ and $D\bar{D}\to\pi\pi$
amplitudes. Since the particles involved are all pseudoscalar
particles, we can expand the amplitudes
as \begin{align}
\mathcal{M}_{D\bar{D}} = \sum_{J} (2J+1) P_J(\cos\theta)
M_J(t,\,\cos\theta),\;\;\; \mathcal{M}_{D\bar{D}\to\pi\pi} = \sum_{J}
     (2J+1) P_J(\cos\theta) A_J(t,\,\cos\theta).
\end{align}
Thus, we obtain the spectral functions for the $D\bar{D}$ amplitude
\begin{align}
\rho_{JT}^{(\pm)} (t) = \mathrm{Im}   M_{JT}^{D\bar{D}} =
\frac1{32\pi} \sqrt{\frac{t-4m_\pi^2}{t}}
|A_{JT}^{D\bar{D}\to\pi\pi} |^2 .
\label{eq:spect}
\end{align}
Note that we have to subtract the Born amplitudes from
Eq.~\eqref{eq:spect}, i.e., the spectral function is in fact defined
as
\begin{align}
 \rho_{00}^{(+)}(t) = \mathrm{Im}   M_{00}^{D\bar{D}} -
  \mathrm{Im}   M_{00}^{D\bar{D}, \mathrm{Born}},\;\;\;
 \rho_{11}^{(-)}(t) = \mathrm{Im}  M_{11}^{D\bar{D}} -
  \mathrm{Im}   M_{11}^{D\bar{D}, \mathrm{Born}}.
\label{eq:spect2}
\end{align}
Using the dispersion relation, we find the $DD$ amplitude with
correlated $2\pi$ exchange as
\begin{align}
\mathcal{M}_{DD}^{S-\mathrm{wave\;\;corr.} 2\pi} \;=\;
  \frac1{\pi}\int_{4m_\pi^2}^\infty 
\frac{\rho_{00}^{(+)}(t')}{t-t'}\,dt',\;\;\;
\mathcal{M}_{DD}^{P-\mathrm{wave\,\,corr.} 2\pi} \;=\;
  \frac1{\pi}\int_{4m_\pi^2}^\infty 
\frac{\rho_{11}^{(-)}(t')}{t-t'}\,dt',
\label{eq:final}
\end{align}
as shown in Fig.~\ref{fig:1}. 
Here, we have suppressed the spin structure for the vector-isovector
($\rho$) channel.
Since $D^*$ is a vector meson, the partial-wave expansions of the
$D^*\bar{D}^*\to \pi\pi$ and $D^*\bar{D}^*$ are more involved because
of the spin. Thus, we present the detailed calculation of deriving the
spectral functions for the $D^*\bar{D}^*$ amplitudes in
Appendix~\ref{app:a}. The explicit expressions of the spectral
functions for the $D^*D^*$ channel are given as follows:
\begin{align}
\rho_{00}^{(+),1}(t) &=
\frac{3}{4M_{D^*}^2}\left[\mathrm{Im}\,p_1^{+,J=0}(t) -
  \mathrm{Im}\,p_{1,\mathrm{Born}}^{+,J=0}(t)\right],\cr
\rho_{11}^{(-),1}(t) &=
 \frac{2}{(4M_{D^*}^2-t)}\left[\mathrm{Im}\,p_1^{-,J=1}(t)  -
  \mathrm{Im}\,p_{1,\mathrm{Born}}^{-,J=1}(t)\right],\cr
\rho_{11}^{(-),2}(t) &= \frac{2M_{D^*}^2}{t(4M_{D^*}^2-t)}
\left[\mathrm{Im}\,p_2^{-,J=1}(t) -
  \mathrm{Im}\,p_{2,\mathrm{Born}}^{-,J=1}(t)\right], \cr
\rho_{11}^{(-),3}(t) &=  \frac{2M_{D^*}}{4\sqrt{t}(4M_{D^*}^2-t)}
\left[\mathrm{Im}\,p_3^{-,J=1}(t) -
  \mathrm{Im}\,p_{3,\mathrm{Born}}^{-,J=1}(t)\right],
\label{eq:spectDstar}
\end{align}
where the definitions of $p_i^{\pm}$ can be found in
Appendix~\ref{app:a}. The $D^*D^*$ amplitudes with correlated $2\pi$ 
exchange can be obtained by the dispersion relations, which are
similar to Eq.~\eqref{eq:final}.

\subsection{$\sigma$ and $\rho$ coupling constants}
As shown in Eq.~\eqref{eq:final}, we can determine the $DD$ and
$D^*D^*$ amplitudes with correlated $2\pi$ exchange. On the other
hand, it is difficult to extract the $\sigma$ and $\rho$ coupling
constants without any approximations. The best way to determine the
coupling constants is first to compute the $DD$ and
$D^*D^*$ amplitudes by using the effective Lagrangians, and then
compare them to those with correlated $2\pi$ exchange. Thus, we will
first derive the $DD$ and $D^*D^*$ amplitudes based on the
following effective Lagrangians:
\begin{align}
\mathcal{L}_{DD\sigma} &= 2g_{DD\sigma} M_D DD^\dagger \sigma,\cr
\mathcal{L}_{DD\rho} &= ig_{DD\rho}
\left( D \bm{\tau}\cdot\bm{\rho}^\mu \partial_\mu D^\dagger
- D^\dagger \bm{\tau}\cdot\bm{\rho}^\mu \partial_\mu D \right),\cr
\mathcal{L}_{D^*D^*\sigma}&=2M_{D^*} g_{D^*D^*\sigma}
D^{*\mu}\bar{D}_{\mu}^*\sigma,\cr
\mathcal{L}_{D^*D^*\rho} &= i g_{D^*D^*\rho}
(\bar{D}^{*\nu} \bm{\tau} \cdot \bm{\rho}^\mu \partial_\mu D_{\nu}^*
- D_\nu^* \bm{\tau} \cdot \bm{\rho}^\mu \partial_\mu \bar{D}_\nu^*)
+4if_{D^*D^*\rho} \bar{D}^*_\mu \bm{\tau}  \cdot
(\partial^\mu \bm{\rho}^\nu - \partial^\nu \bm{\rho}^\mu) D^*_\nu,
\label{eq:effLag2}
\end{align}
where $\sigma$ and $\bm{\rho}_\mu$ denote the $\sigma$- and
$\rho$-meson fields. We want to mention that for the $DD\sigma$ and
$D^*D^*\sigma$ Lagrangians, we need to introduce additional
dimensionful parameters, i.e., the masses of the $D$ and $D^*$ mesons,
respectively. Here, the $D$ and $D^*$ mesons are not the scaled
fields, which are different from $P$ and $P^*$ fields by the scaling
factors $(M_{D})^{-1/2}$ and $(M_{D^*})^{-1/2}$, respectively. As for
the $D^*D^*\rho$ vertices, we have two different coupling constants,
i.e., the vector coupling constant $g_{D^*D^*\rho}$ and the tensor
coupling constant $f_{D^*D^*\rho}$, which will be also determined in
the present work.

Using the effective Lagrangians in Eq.~\eqref{eq:effLag2},
we obtain the invariant amplitudes for the $DD\to DD$ and $D^*D^*\to
D^*D^*$ processes as follows:
\begin{align}
\mathcal{M}_{DD}^{\sigma}(t) &= g_{DD\sigma}^2
\frac{4M_D^2}{t-m_\sigma^2},  \;\;\;
\mathcal{M}_{DD}^{\rho}(t,s) = g_{DD\rho}^2 \frac{s-u}{t-m_\rho^2},\cr
\mathcal{M}_{D^*D^*}^{\sigma,\lambda_1,\lambda_2,\lambda_3,\lambda_4}(t,s)&= 16
\epsilon_\mu(\mathbf{p},\lambda_1) \epsilon_\nu(-\mathbf{p},\lambda_2)
\epsilon_{\alpha}^*(\mathbf{p'},\lambda_3)
 \epsilon_{\beta}^*(-\mathbf{p}',\lambda_4) \frac{g_{D^*D^*\sigma}^2
 M_{D^*}^2}{t-m_\sigma^2} \mathcal{A}^{\mu \nu \alpha \beta},\cr
\mathcal{M}_{D^*D^*}^{\rho,\lambda_1,\lambda_2,\lambda_3,\lambda_4}
  (t,s) &=
\epsilon_\mu(\mathbf{p},\lambda_1) \epsilon_\nu(-\mathbf{p},\lambda_2)
\left\{ 4g_{D^*D^*\rho}^2 \frac{s-u}{t-m_\rho^2} \mathcal{A}^{\mu \nu \alpha \beta}
 + 32\sqrt{3} f_{D^*D^*\rho}^2 \frac{t}{t-m_\rho^2}
\mathcal{B}^{\mu \nu \alpha \beta}\right. \cr
&\left. +\,16g_{D^*D^*\rho}f_{D^*D^*\rho}
  \frac{\sqrt{t(4M_{D^*}^2-t)}}{t-m_\rho^2}
\mathcal{C}^{\mu \nu \alpha \beta} \right\}
\epsilon_{\alpha}^*(\mathbf{p'},\lambda_3)
 \epsilon_{\beta}^*(-\mathbf{p}',\lambda_4),
\label{eq:transamp}                                                
\end{align}
where $\lambda_i$ stand for the helicities of the corresponding
$D^*$ mesons in both the initial and final states.
$\mathcal{A}^{\mu\nu\alpha\beta}$, $\mathcal{B}^{\mu\nu\alpha\beta}$ and
$\mathcal{C}^{\mu\nu\alpha\beta}$ denote the projection operators,
which can be also found in Appendix~\ref{app:a}.

While the spectral functions we have explicitly derived in the present work
as shown in Eqs.~\eqref{eq:spect2} and~\eqref{eq:spectDstar} contain
information on the coupling strength for the $\sigma$ and $\rho$
mesons, it is rather difficult to extract the exact values of them.
One possible way of extracting the coupling constants from the spectral
functions is to make a pole approximation that is expressed, for
example, by 
\begin{align}
\rho_{00}^{(+)}(t') = \pi g_{DD\sigma}^2 \delta(t'-m_\sigma^2), \;\;\;
  \rho_{11}^{(-)} (t') =  \pi g_{DD\rho}^2 \delta(t'-m_\rho^2),
\label{eq:pole_approx}  
\end{align}
where $g_{DD\sigma}$ and $g_{DD\rho}$ denote the \emph{on-mass-shell}
coupling constants for the $DD\sigma$ and $DD\rho$ vertices,
respectively. These on-mass-shell coupling constants are used for the
description of $DD$ or $D\bar{D}$ reactions. 
Then we are able to reproduce all the amplitudes obtained from the
effective Lagrangians such as 
\begin{align}
\frac{1}{\pi}
  \int_{4m_\pi^2}^\infty \frac{\rho_{00}^{(+)} (t')}{t-t'} \,dt'
  \approx \frac{g_{DD\sigma^2}}{t-m_{\sigma}^2},\;\;\; t\le 0. 
\end{align}
We can apply the same pole approximations to the $D^*D^*$ case.  Thus,  
the on-mass-shell coupling constants can be written by 
\begin{align}
g_{DD\sigma}^2 &\approx  \frac{t-m_\sigma^2}{\pi}
                 \int_{4m_\pi^2}^\infty  
\frac{\rho_{00}^{(+)}(t')dt'}{t-t'},\;\;\;
g_{DD\rho}^2 \approx \frac{t -m_\rho^2}{\pi} \int_{4m_\pi^2}^\infty
\frac{\rho_{11}^{(-)}(t')dt'}{t-t'},\cr
g_{D^*D^*\sigma}^2 &\approx \frac{t-m_\sigma^2}{\pi}
\int_{4m_\pi^2}^\infty
 \frac{\rho_{00}^{(+),1}(t')dt'}{t-t'},\;\;\;
g_{D^*D^*\rho}^2 \approx \frac{t-m_\rho^2}{\pi}
\int_{4m_\pi^2}^\infty \frac{\rho_{11}^{(-),1}(t')dt'}{t-t'},\cr
f_{D^*D^*\rho}^2 &\approx \frac{t-m_\rho^2}{\pi}
\int_{4m_\pi^2}^\infty \frac{\rho_{11}^{(-),2}(t')dt'}{t-t'},\;\;\;
f_{D^*D^*\rho} g_{D^*D^*\rho} \approx \frac{t-m_\rho^2}{\pi}
\int_{4m_\pi^2}^\infty \frac{\rho_{11}^{(-),3}(t')dt'}{t-t'},
\label{eq:couplings}
\end{align}
where $t$ is the square of the momentum transfer in the $s$ channel,
i.e., $t\le 0$. Note that the left-hand sides of the above expressions 
contain the $t$ variable. However, the approximated one-mass-shell
coupling constants are almost independent of $t$. Only the numerical
result of the $D^*D^*\sigma$ coupling constant exhibits mild
dependence on $t$, which comes from the broad width of the $\sigma$
meson as implied in Eq.~\eqref{eq:couplings}.

On the other hand, various works including lattice
QCD~\cite{Can:2012tx, Bracco:2001dj, El-Bennich:2016bno}  derive 
the coupling constants for the $\rho$ meson not on the corresponding
mass shell but at $t=0$. Actually, the coupling constant at $t=0$
reflects the effect from a form factor that reduces the coupling
strength by approximately a difference between the square of the
cutoff mass and the mass of the corresponding exchanged meson. When    
exotic heavy mesons such as the $X$, $Y$, and $Z$ mesons are 
investigated in a meson-exchange picture, 
a monopole-type form factor is often used~\cite{Liu:2008tn, Lee:2009hy,
Liu:2010xh, Liu:2019stu}. The transition amplitude for $\sigma$
exchange is expressed as  
\begin{align}
\mathcal{T}_{DD}^\sigma(t) =
  \frac{g_{DD\sigma}^2}{t-m_\sigma^2}\left(\frac{\Lambda_{\sigma}^2 
  -m_\sigma^2}{\Lambda_{\sigma}^2 - t} \right)^2.
  \label{eq:sigma_amp}
\end{align}
If we take into account Eq.~\eqref{eq:sigma_amp} and the pole
approximation given in Eq.~\eqref{eq:pole_approx}, we are able to
write a phenomenological expression for the vertex function
$g_{DD\sigma}(t)$ 
\begin{align}
  \label{eq:corre2pi_ff}
g_{DD\sigma}^2(t)  = \frac{t-m_\sigma^2}{\pi}
  \int_{4m_\pi^2}^\infty \frac{\rho_{00}^{(+)} (t')}{t-t'}
  \left(\frac{\Lambda_\sigma^2 -   t'}{\Lambda_\sigma^2
  -t}\right)^2\,dt',\;\;\; t\le 0,
\end{align}
where we have introduced a $t'$-dependent form factor
\begin{align}
F(t,t') = \frac{\Lambda_\sigma^2 -   t'}{\Lambda_\sigma^2  -t} .
\end{align}
A similar expression was used in the $NN$
interaction~\cite{Kim:1993xw}.  The other vertex functions for
the $DD\rho$, $D^*D^*\sigma$, and $D^*D^* \rho$ vertices can be
written in a similar way. Using Eq.~\eqref{eq:corre2pi_ff}, we can
compare the present results of the \emph{off-mass-shell} coupling
constants with those from other works at least phenomenologically.      

Since the $\sigma$ meson has a very broad mass, one has to examine the
dependence of the coupling constants on the mass of the $\sigma$
meson. Note that when we perform the integrals in
Eq.~\eqref{eq:couplings} we take the upper limit to be $52m_\pi^2$,
which was usually done in the case of the $NN$ interaction. In fact,
it is well known that the contributions from the spectral functions at 
higher $t'$ are rather small, when one considers the $\sigma$ and
$\rho$ meson channels. As menioned in Introduction, note that we have
not included the $K\bar{K}$ channel. In principle, we can introduce
it, utilizing the coupled channel formalism. However, since we do not
know the coupling constants for the $D_sD^* K$, $DD_s^*K$ and
$D^*D_s^*K$ vertices both experimentally and theoretically, we have to
consider these coupling constants as free parameters, which brings
about unavoidably additional uncertainties in the present work. Thus,
we will take into account only the $\pi\pi$ channel. 

\section{Results and Discussion}
To compute first the off-shell $D\bar{D}\to \pi\pi$ amplitudes in the
pseudophysical region, we need to determine the coupling constants for
the $DD^*\pi$ and $D^*D^*\pi$ vertices. we use the value of the
$g_{DD^*\pi}$ determined by the CLEO Collaboration~\cite{Ahmed:2001xc,
Anastassov:2001cw}. The $g_{DD^*\pi}$ and $g_{D^*D^*\pi}$ are
related to the coupling $g$ in the effective Lagrangians in
Eq.~\eqref{eq:Lagrangian} as follows: 
\begin{align}
g_{DD^*\pi} = \frac{2 g}{f_\pi} \sqrt{M_D M_{D^*}},\;\;\;
g_{D^*D^*\pi} = \frac{2 g}{f_\pi},
\end{align}
which can be found by using the decay rate of the $D^*$ meson. The
strong coupling $g$ is known to be $g=0.59$. We will use in this work
the value determined by the CLEO Collaboration
$g_{DD^*\pi}=17.9$. If one considers the mass difference between the
$D$ and $D^*$ mesons, $g_{DD^*\pi}$ would become 17.3. As we have
already discussed in the previous Section, we take the 
numerical values of the cutoff masses as $\Lambda_{D^*}=2.8$ GeV and
$\Lambda_D=2.5$ GeV. The results depend marginally on the values of
the cutoff masses. The uncertainty, which arises from the cutoff
masses, is about $20\,\%$. Then we can proceed to compute numerically
the spectral functions for the $DD$ and $D^*D^*$ amplitudes with
correlated $2\pi$ exchange, which are expressed in
Eqs.~\eqref{eq:spect2} and~\eqref{eq:spectDstar}.

\begin{figure}[htp]
\centering
\includegraphics[scale=0.23]{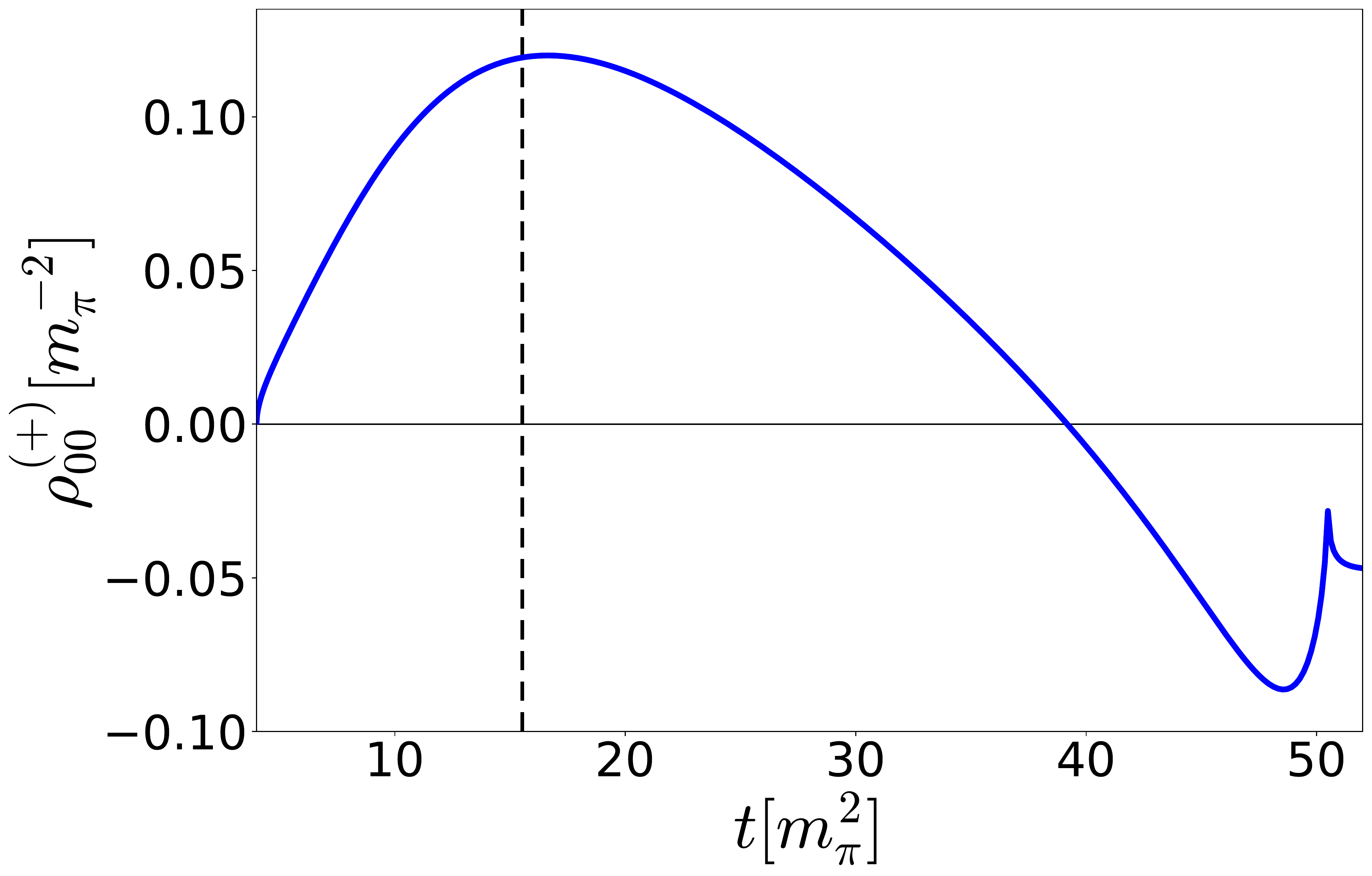}
\includegraphics[scale=0.23]{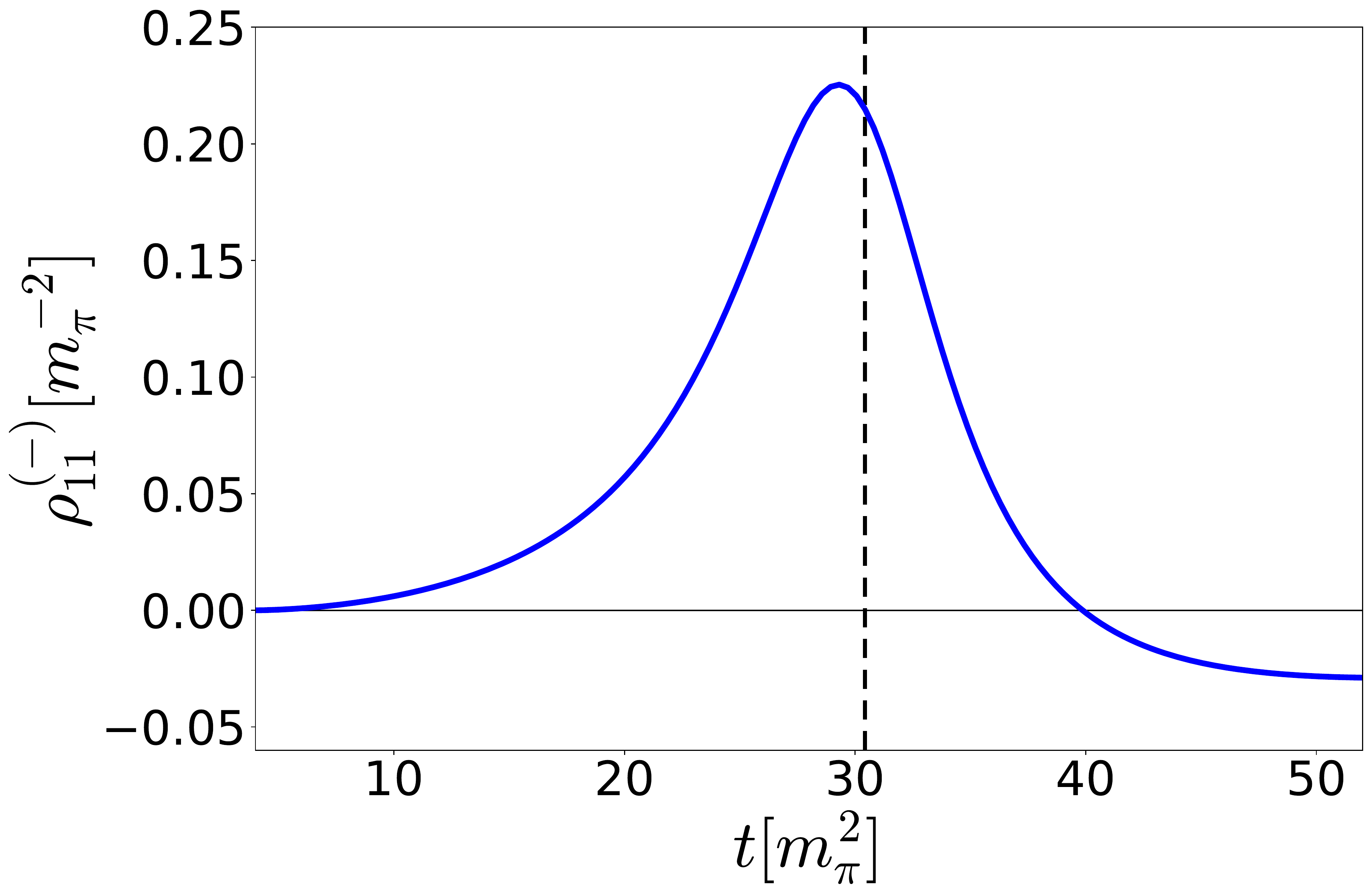}
  \caption{The spectral function $\rho_{00}^{(+)}$ (in the right
    panel) and $\rho_{11}^{(-)}$ for the $D\bar{D}$
    channel as a function of $t$ in unit of $m_\pi^{-2}$. Note that
    $t$ denotes the square of the center-of-mass energy in the $t$
    channel. The dashed vertical line in the left panel corresponds to
    the $\sigma$-meson mass $550$ MeV, whereas that in the right panel 
    designates the $\rho$-meson mass $770$ MeV.}
  \label{fig:5}
\end{figure}
In the left panel of Fig.~\ref{fig:5}, we draw the result of the
spectral function $\rho_{00}^{(+)}$, of which the expression is given in
Eq.~\eqref{eq:spect2}. Its broad shape arises from the resonance of
the $\sigma$ meson with the large width. We see that the spectral
function falls off after around $t=18m_\pi^2$ and then becomes
negative from $t=30m_\pi^2$, which is similar to the case of the
$N\bar{N}$ interaction~\cite{Kim:1994ce}. Since the width of the
$\sigma$ meson is rather large, one should consider the dependence of
the $DD\sigma$ coupling constant on the mass of the $\sigma$ meson,
$m_\sigma$, which will be explicitly shown later.
The right panel of Fig.~\ref{fig:5} shows the result of
$\rho_{11}^{(-)}$ given in Eq.~\eqref{eq:spect2}, which yields the
$g_{DD\rho}$ coupling constant. We observe that $\rho_{11}^{(-)}$
becomes negative from around $t=40m_\pi^2$, which is again similar to
the $N\bar{N}$ case in the $\rho$ channel~\cite{Kim:1994ce}.

\begin{figure}[htp]
\centering
\includegraphics[scale=0.23]{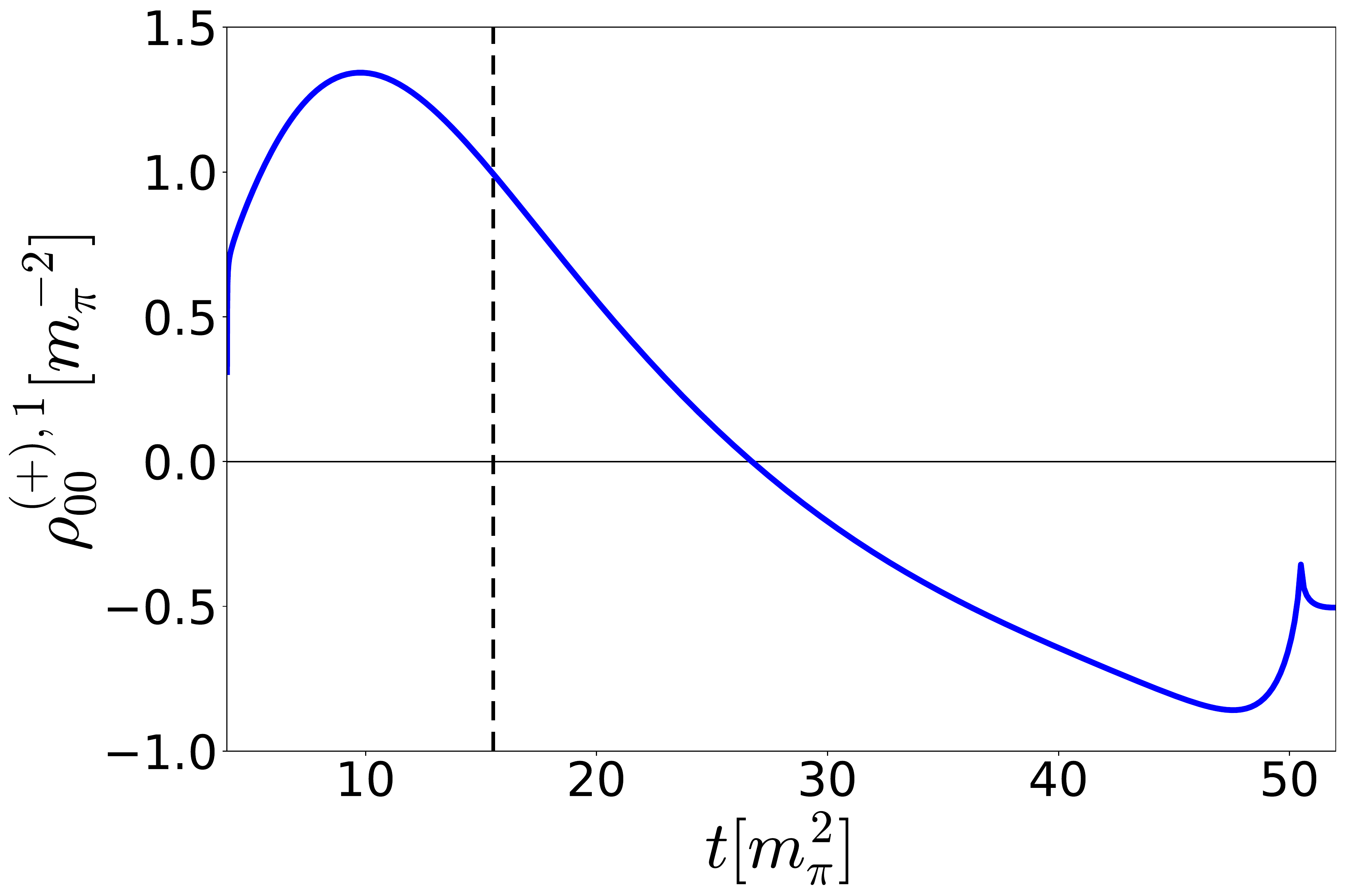}
\includegraphics[scale=0.23]{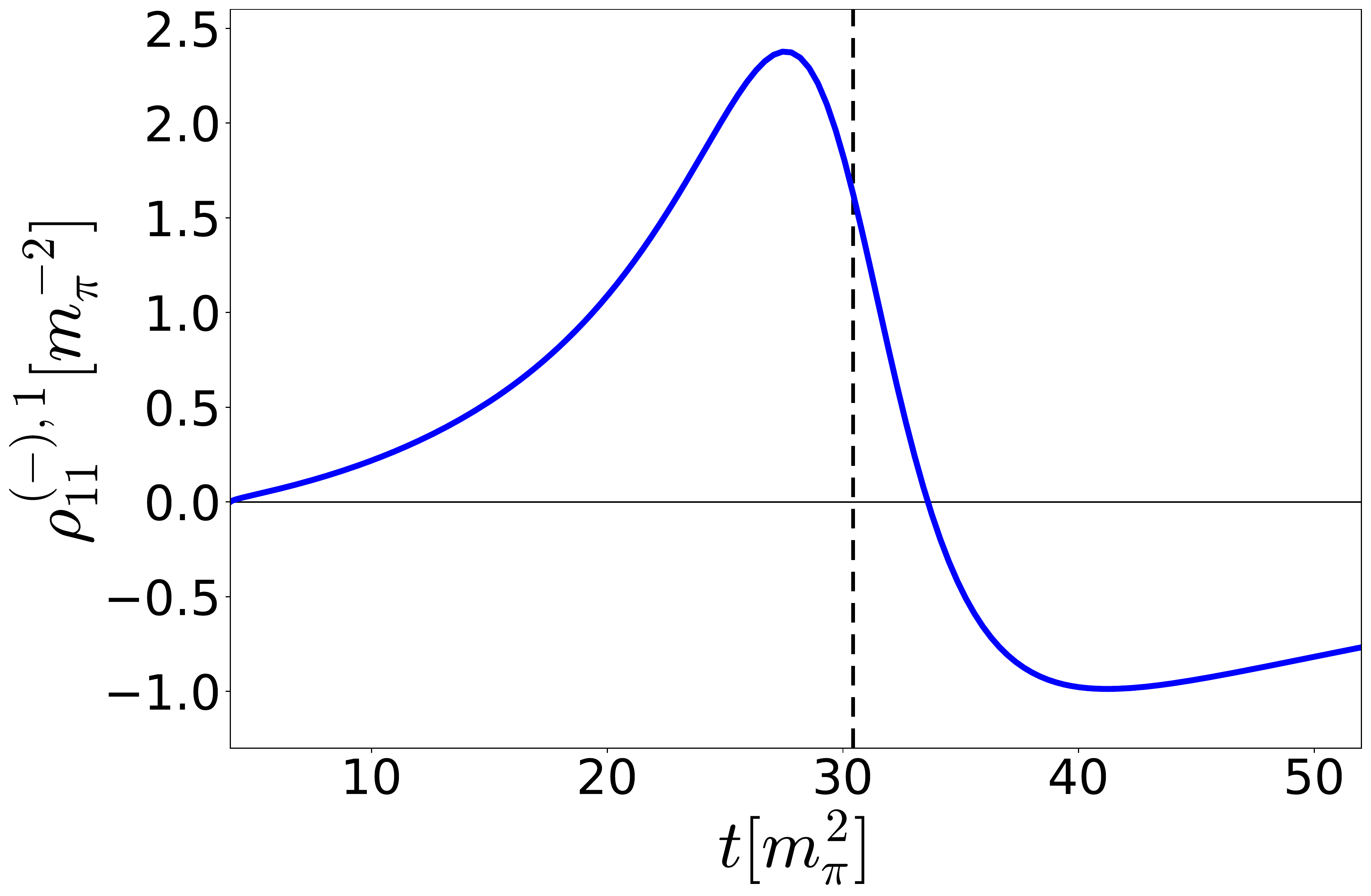}
  \caption{The spectral function $\rho_{00}^{(+),1}$ (left panel) and
    $\rho_{11}^{(-),1}$ (right panel) for the $D^*\bar{D}^*$ channel
    as a function of $t$ in unit of $m_\pi^{-2}$. Notations
    are the same as in Fig.~\ref{fig:5}.}
  \label{fig:6}
\end{figure}
In the left panel of Fig.~\ref{fig:6}, we depict the result of the
spectral function for the $D^*\bar{D}^*$ case in the $\sigma$
channel. It looks different from the corresponding $D\bar{D}$
case. The reason is that the Born amplitudes for the $D^*\bar{D}^*\to
\pi\pi$ have rather strong dependence on $t$. Since we subtract the
modulus squared of the Born amplitudes to avoid any
double countings that arise from the whole $2\pi$ exchange (see, for
example, Eq.~\eqref{eq:spect2}), the $D^*\bar{D}^*$ spectral function
$\rho_{00}^{(+),1}$ becomes negative already from around $28m_\pi^2$.
In the right panel, we draw the result of $\rho_{11}^{(-),1}$, which
is defined in Eq.~\eqref{eq:spectDstar}. It will provide the vector
coupling constant $g_{D^*D^*\rho}$.
\begin{figure}[ht]
\includegraphics[scale=0.23]{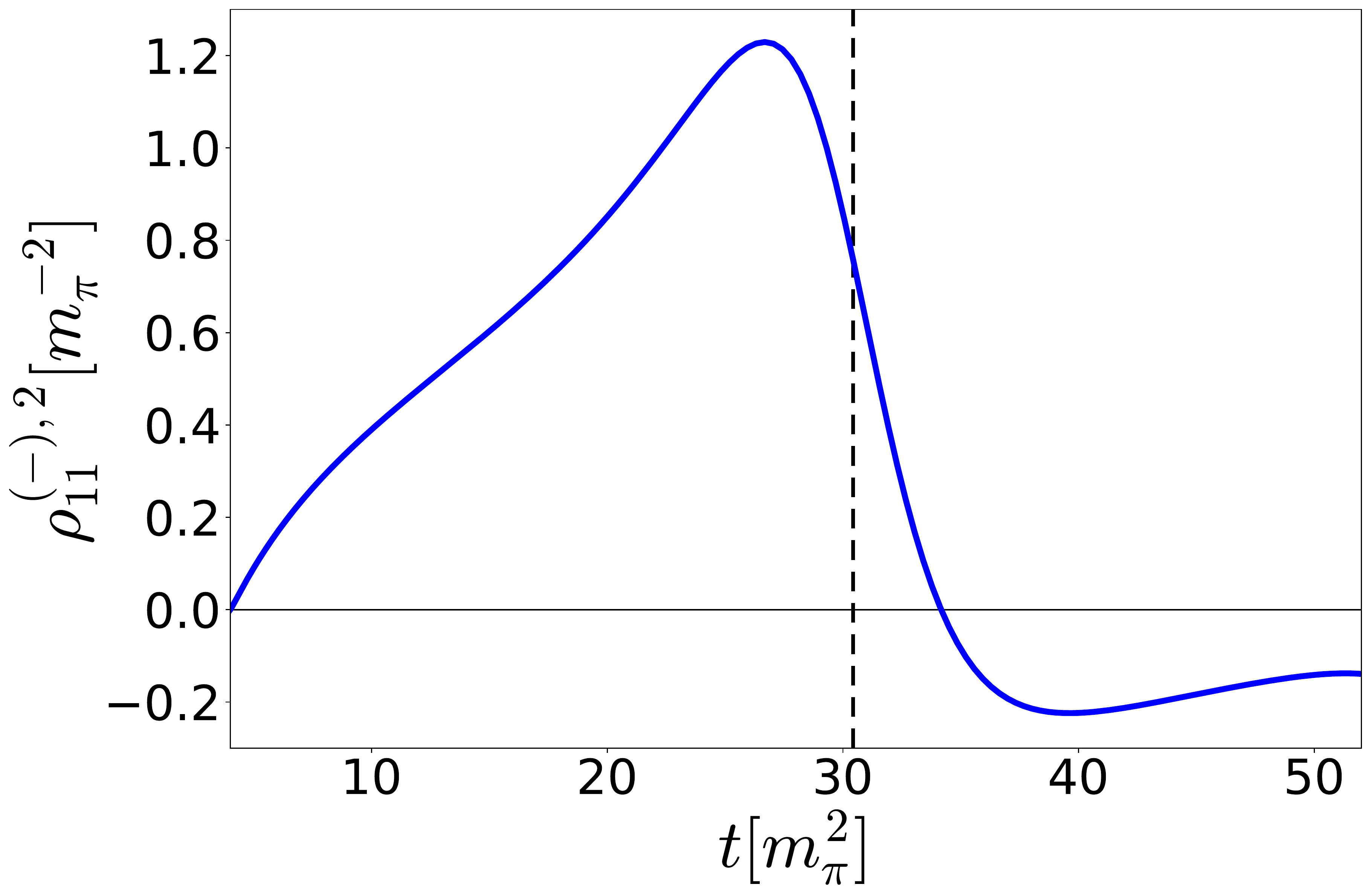}
\includegraphics[scale=0.23]{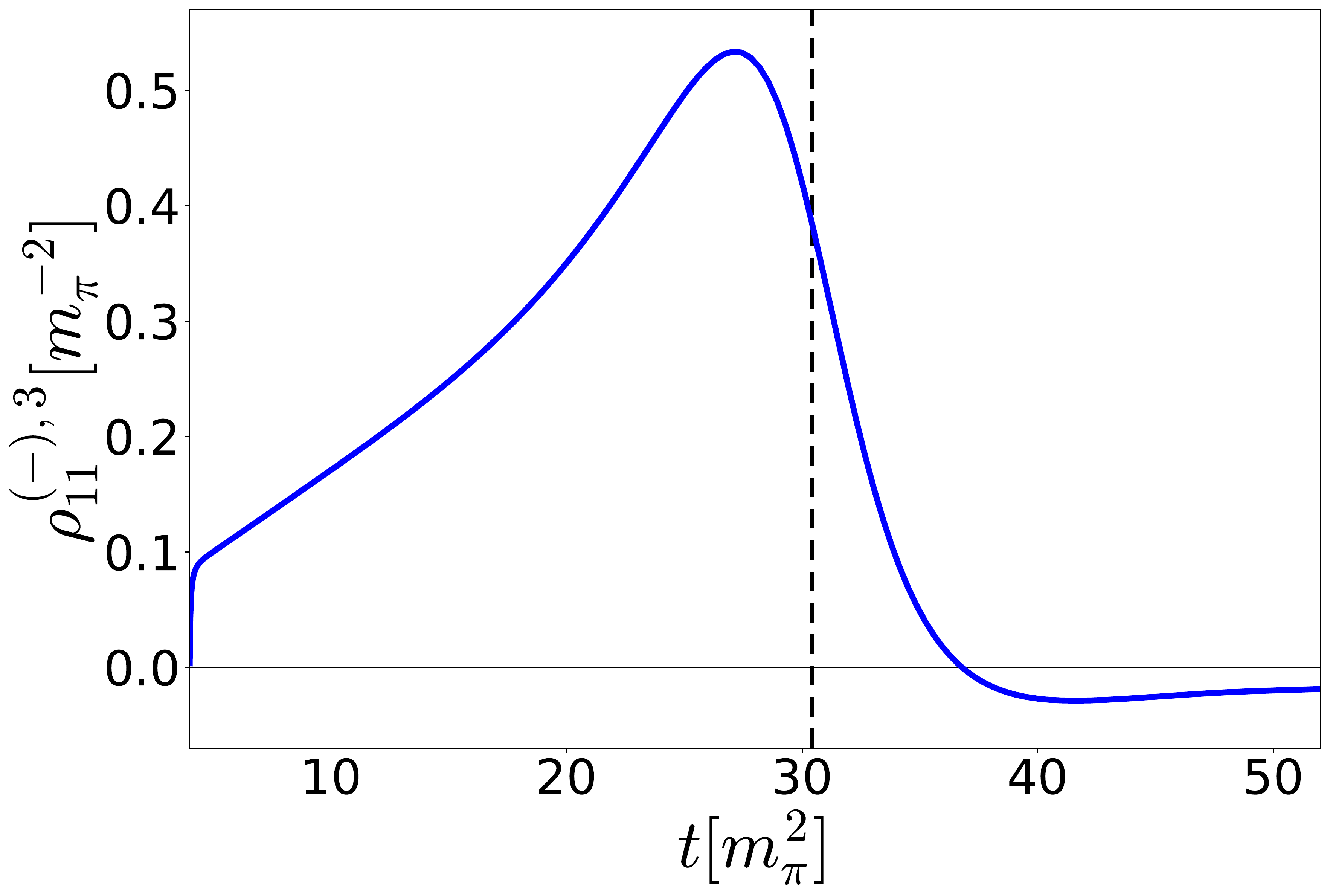}
  \caption{The spectral functions $\rho_{11}^{(-),2}$ (left panel) and
    $\rho_{11}^{(-),3}$ (right panel) for the $D^*\bar{D}^*$ channel
    as a function of $t$ in unit of $m_\pi^{-2}$. Notations are the
    same as in Fig.~\ref{fig:5}.}
  \label{fig:7}
\end{figure}
In the left and right panels of Fig.~\ref{fig:7}, we present the
numerical results of the spectral functions $\rho_{11}^{(-),2}$ and
$\rho_{11}^{(-),3}$, respectively. They are less affected by the
subtraction of the Born terms, compared to the results of
$\rho_{11}^{(-),1}$. 

\begin{figure}[ht]
\includegraphics[scale=0.52]{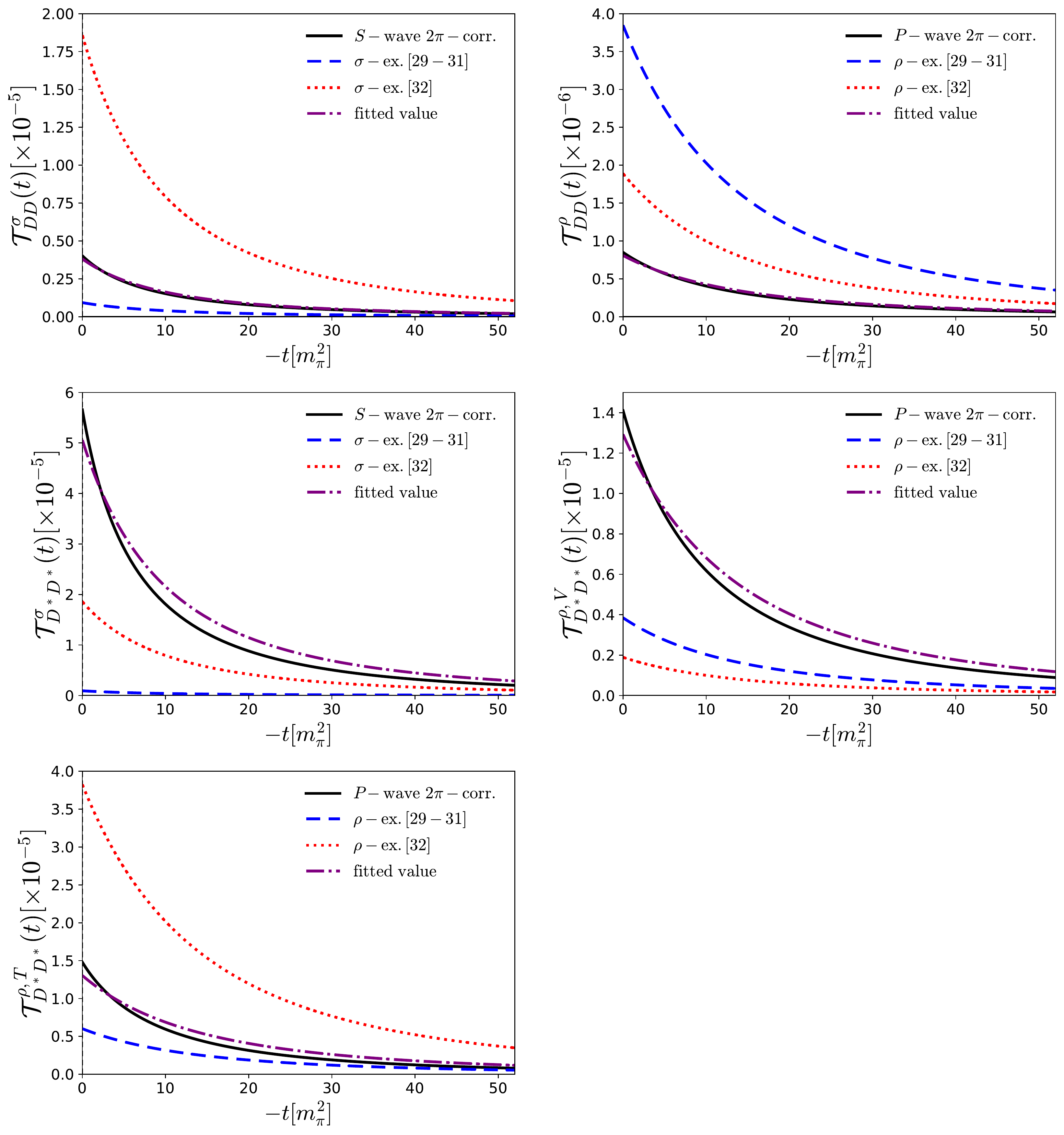}
\caption{ The $DD$ and $D^*D^*$ amplitudes with $S$- and $P$-wave
    correlated $2\pi$ exchange contributions, respectively. Solid
    curves depict the amplitudes with correlated $2\pi$ exchange
    whereas dashed and dotted ones show the amplitudes produced by
    using the coupling constants taken from Refs.~\cite{Liu:2008tn,
      Lee:2009hy, Liu:2010xh} and Ref.~\cite{Liu:2019stu},
    respectively.
Dash-dotted curves are obtained by fitting the $\sigma$ and $\rho$
coupling constants in such a way that the amplitudes with $\sigma$ and
$\rho$ exchanges reproduce those with $S$- and $P$-wave
    correlated $2\pi$ exchange, respectively. 
  }
  \label{fig:8}
\end{figure}
Before we extract the $\sigma$ and $\rho$ coupling constants, we 
first examine the transition amplitudes for $DD\to DD$ and $D^*D^* \to
D^*D^*$ with $S$- and $P$-wave correlated $2\pi$ exchange
contributions, respectively. As we have mentioned already, 
the transition amplitudes for the $DD$ and $D^*D^*$ 
processes with $\sigma$ and $\rho$ exchanges given explicitly in
Eq.~\eqref{eq:transamp} are compatible with those with 
$S$- and $P$-wave correlated $2\pi$ exchanges, which are shown in
Eq.~\eqref{eq:final}. We use the values of the cutoff masses
$\Lambda_{\sigma({\rho})} =1$ GeV, which is used in other
works~\cite{Liu:2008tn, Lee:2009hy, Liu:2010xh, Liu:2019stu}.  
In Fig.~\ref{fig:8} we draw the results for the $DD$ and $D^*D^*$
transition amplitudes with $S$- and $P$-wave 
correlated $2\pi$ exchange contributions in
Eq.~\eqref{eq:corre2pi_ff}. In the upper left panel, the present
result for the $DD$ amplitude with $S$-wave 
correlated $2\pi$ exchange is depicted in the solid curve, compared
with those with $\sigma$ exchange as given in
Eq.~\eqref{eq:sigma_amp}, for which the values of the $DD\sigma$  
coupling constant are taken from Refs.~\cite{Liu:2008tn, Lee:2009hy,
  Liu:2010xh} and Ref.~\cite{Liu:2019stu}, respectively shown in the
dashed and dotted curves.

The best way to determine the \emph{on-mass-shell} $\sigma$ and $\rho$
coupling constants is to fit the $DD$ and $D^*D^*$ amplitudes
presented in Fig.~\ref{fig:8} by changing these coupling
constants. Dot-dashed curves in Fig.~\ref{fig:8} are extracted by
fitting the $\sigma$ and $\rho$ coupling constants such that the
$\sigma$- and $\rho$-exchange amplitudes reproduce the $DD$ and
$D^*D^*$ amplitudes. As shown in the upper panels of Fig.~\ref{fig:8},
we are able to reproduce very well the $DD$ amplitudes with $S$- and
$P$-wave correlated $2 \pi$ exchanges. The $DD\sigma$ and $DD\rho$
coupling constants are determined to be $g_{DD\sigma}=1.50$ and
$g_{DD\rho}=1.65$, respectively. The present result lies between that
with $g_{DD\sigma}=0.76$~\cite{Liu:2008tn, Lee:2009hy, Liu:2010xh} and
that with $g_{DD\sigma}=3.4$~\cite{Liu:2019stu}. 
This implies that the extracted $g_{DD\sigma}$ coupling constant from
the present work should be found between these two values. On the
other hand, the result for the $DD$ amplitude with $P$-wave correlated
$2\pi$ exchange, which is drawn in the upper right panel of
Fig.~\ref{fig:8}, is much weaker than those with the values of
$g_{DD\rho} =3.71$ and $2.6$, taken respectively from
Refs.~\cite{Liu:2008tn, Lee:2009hy, Liu:2010xh} and
Ref.~\cite{Liu:2019stu}. 

The middle left panel of Fig.~\ref{fig:8} shows that the present
result for the $D^*D^*$ amplitude with $S$-wave correlated $2\pi$
exchange is greater than the other two results for those with
$g_{D^*D^*\sigma}=0.76$ and $3.4$. The present results for the $DD$
and $D^*D^*$ amplitudes already indicate that $g_{DD\sigma}$ and
$g_{D^*D^*\sigma}$ coupling constants are quite different. So far,
many theoretical works on  
heavy meson interactions set these two $\sigma$ coupling constants
equal each other. However, if one considers correlated $2\pi$
exchange, there is at least one reason why $g_{DD\sigma}$ should be
different from $g_{D^*D^*\sigma}$: While the $D\bar{D}\to \pi\pi$
amplitude contains only $D^*$ exchange, the $D^*\bar{D}^* \to 2\pi$
amplitude has both $D$ and $D^*$ exchange. Thus, $g_{D^*D^*\sigma}$
should be naturally greater than $g_{DD\sigma}$ within the present
framework. To determine the $D^*D^*\sigma$ coupling constant, we
again fit the $D^*D^*$ amplitude with $S$-wave correlated $2\pi$
exchange as done in the case of the $DD\sigma$ coupling constant. The
dot-dashed curve in the middle left panel of Fig.~\ref{fig:8} depicts
the numerical result for the $D^*D^*$ amplitude  with 
$g_{D^*D^*\sigma}=5.21$, which describes well that with $S$-wave
correlated $2\pi$ exchange.  This result indicates two important
points: Firstly, the present result for the $D^*D^*\sigma$ coupling
constant is almost 3.5 times larger than that for
$g_{DD\sigma}$. Secondly, the numerical value of $g_{D^*D^*\sigma}$ is
also much larger than those used in the other works ~\cite{Liu:2008tn,
Lee:2009hy, Liu:2010xh, Liu:2019stu}.  

In the middle right and lower left panels of Fig.~\ref{fig:8}
illustrate the present results for the $D^*D^*$ 
amplitudes with vector and tensor $P$-wave correlated $2\pi$ exchange,
respectively. The comparisons of the present results to those with
$\rho$ meson exchange indicate 
that the vector $D^*D^*\rho$ coupling constant extracted from this
work should be larger than $g_{D^*D^* \rho}$ of
Refs.~\cite{Liu:2008tn, Lee:2009hy, Liu:2010xh, Liu:2019stu} whereas
the value of the tensor coupling $f_{D^*D^*\rho}$ from the present
work should lie in between. Indeed, the numerical results for the
vector and tensor $D^*D^*\rho$ coupling constants are evaluated as
follows: $g_{D^*D^*\rho}=6.47$ and $f_{D^*D^*\rho}=6.37$.  The results
are summarized in Table~\ref{tab:1}. 
\begin{table}[ht]
\centering
\caption{$\sigma$ and $\rho$ coupling constant for the $D$ and $D^*$
  mesons. In the second column, the present results are listed. The
  third and fourth ones list the results from Refs.~\cite{Liu:2008tn,
    Lee:2009hy, Liu:2010xh} and Ref.~\cite{Liu:2019stu}, respectively.  
}
\label{tab:1}
\begin{tabular}[t]{c|ccc}
\hline\hline
& Present work & \cite{Liu:2008tn,
Lee:2009hy, Liu:2010xh}   & \cite{Liu:2019stu}\\  
\hline
$g_{DD\sigma}$ & $1.50$ & $0.76$ & $3.4$ \\
$g_{DD\rho}$ & $1.65$ & $3.71$ & $2.6$ \\
$g_{D^*D^*\sigma}$ & $5.21$ & $0.76$ & $3.4$ \\
$g_{D^*D^*\rho}$ & $6.47$ & $3.71$ & $2.6$ \\
$f_{D^*D^*\rho}$ & $6.37$ & $4.64$ & $11.7$ \\
\hline\hline
\end{tabular}
\end{table}

We will now discuss the physical
implications of the present results for the $\sigma$ and $\rho$
coupling constants in comparison with those used in the other works. 
Since there is no information on the $DD\sigma$ coupling constant, the
nonlinear sigma model has been often employed to determine
$g_{DD\sigma}$. Furthermore, the coupling constant $g_{D^*D^*\sigma}$
was naively set equal to $g_{DD\sigma}$, the heavy-quark spin symmetry
being assumed. However, it is well known from the $NN$ interaction
that $\sigma$ exchange arises from a pole 
approximation of $S$-wave (scalar-isoscalar) correlated $2\pi$
exchange~\cite{Brown, Machleidt:1987hj, Kim:1994ce}. Moreover, the
$\sigma$ meson with broad width cannot be identified as the chiral
partner of the pion. This implies that the $\sigma$ coupling constant
can only be quantitatively determined by considering the correlated
$2\pi$ exchange in the scalar-isoscalar channel. Note that
the $\sigma$ coupling constants for any hadrons are not just mere
parameters but very dynamical ones. As shown in Table~\ref{tab:1} and
mentioned already prevously, we find that the values of
$g_{D^*D^*\sigma}$ turn out different from those of
$g_{DD\sigma}$. This can be understood by 
examining Fig.~\ref{fig:3}. $D^*\bar{D}^* \to \pi\pi$ amplitudes
receive contributions both from $D$ and $D^*$ exchange whereas the
$D\bar{D} \to \pi\pi$ amplitudes has only the contribution from $D^*$ 
exchange, as we have already discussed previously. So, the magnitude
of the $D^*\bar{D}^*\to \pi\pi$ amplitudes 
is indeed larger than that of the $D\bar{D} \to \pi\pi$
amplitudes. This indicates that $g_{D^*D^*\sigma}$ should
naturally be larger than $g_{DD\sigma}$. The present result is in
contrast with those of Refs.~\cite{Liu:2008tn,Lee:2009hy,
  Liu:2010xh, Liu:2019stu} and Ref.~\cite{Liu:2019stu}. Furthermore,
there is no consensus on the values of both $g_{DD\sigma}$ and
$g_{D^*D^*\sigma}$. The present result for $g_{DD\sigma}$ is
approximately twice smaller than that of Ref.~\cite{Liu:2019stu},
whereas it is about two times larger than that used in
Ref.~\cite{Liu:2008tn, Lee:2009hy, Liu:2010xh}. On 
the other hand, it is other way around for the value of
$g_{D^*D^*\sigma}$. The present result is approximately seven times 
larger than that of Ref.~\cite{Liu:2008tn, Lee:2009hy, Liu:2010xh},
while it is about 1.5 times larger than that of Ref.~\cite{Liu:2019stu}.  

The $DD\rho$ and $D^*D^*\rho$ coupling constants are usually
determined by using the vector-meson
dominance~\cite{Casalbuoni:1992dx}. Interestingly,
Refs.~\cite{Liu:2008tn, Lee:2009hy, Liu:2010xh, Liu:2019stu} do not
agree on the values of the $\rho$-meson couplings each other, even
though they use the same vector-meson dominance. In particular,  the
values of the tensor coupling constant $f_{D^*D^*\rho}$ differ by
about 2.5 each other. This indicates that no clear consensus exists 
in the values for the $\rho$-meson coupling constants. They do come
again yet from the $P$-wave correlated $2\pi$ exchange in the
vector-isovector channels. The present result for the $DD\rho$
coupling constant is smaller than the results from
Refs.~\cite{Liu:2008tn, Lee:2009hy, Liu:2010xh,Liu:2019stu}. The
present value of the  $D^*D^*\rho$ vector coupling constant is quite larger
than those of Refs.~\cite{Liu:2008tn, Lee:2009hy,
  Liu:2010xh,Liu:2019stu}. On the other hand, that of the tensor
coupling constant $f_{D^*D^*\rho}$ lies between 
that of Refs.~\cite{Liu:2008tn, Lee:2009hy, Liu:2010xh} and that of 
Ref.~\cite{Liu:2019stu}. In particular, it is approximately two times
smaller than that of Ref.~\cite{Liu:2019stu}.

\begin{figure}[htp]
\centering
\includegraphics[scale=0.23]{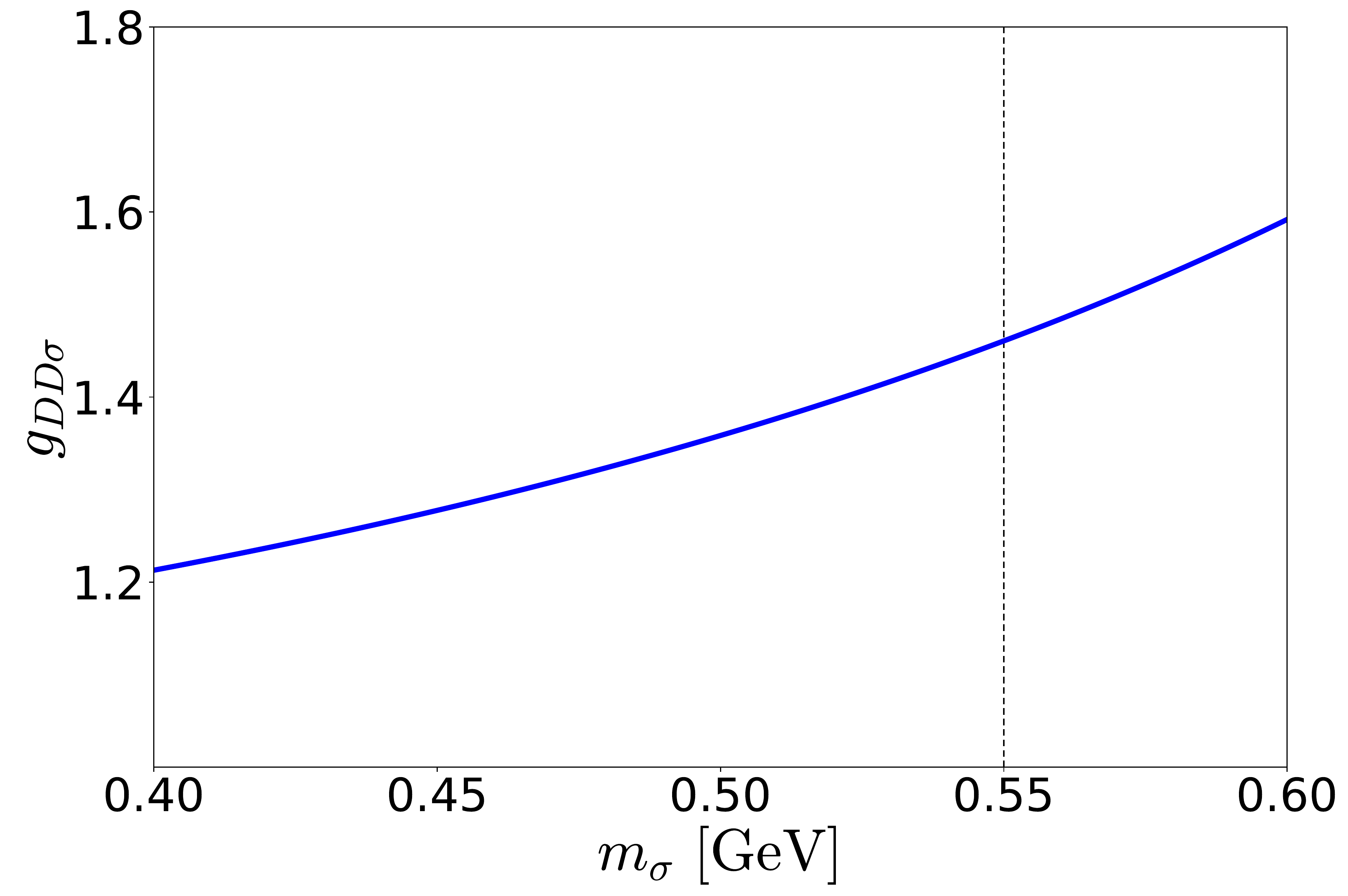}
\includegraphics[scale=0.23]{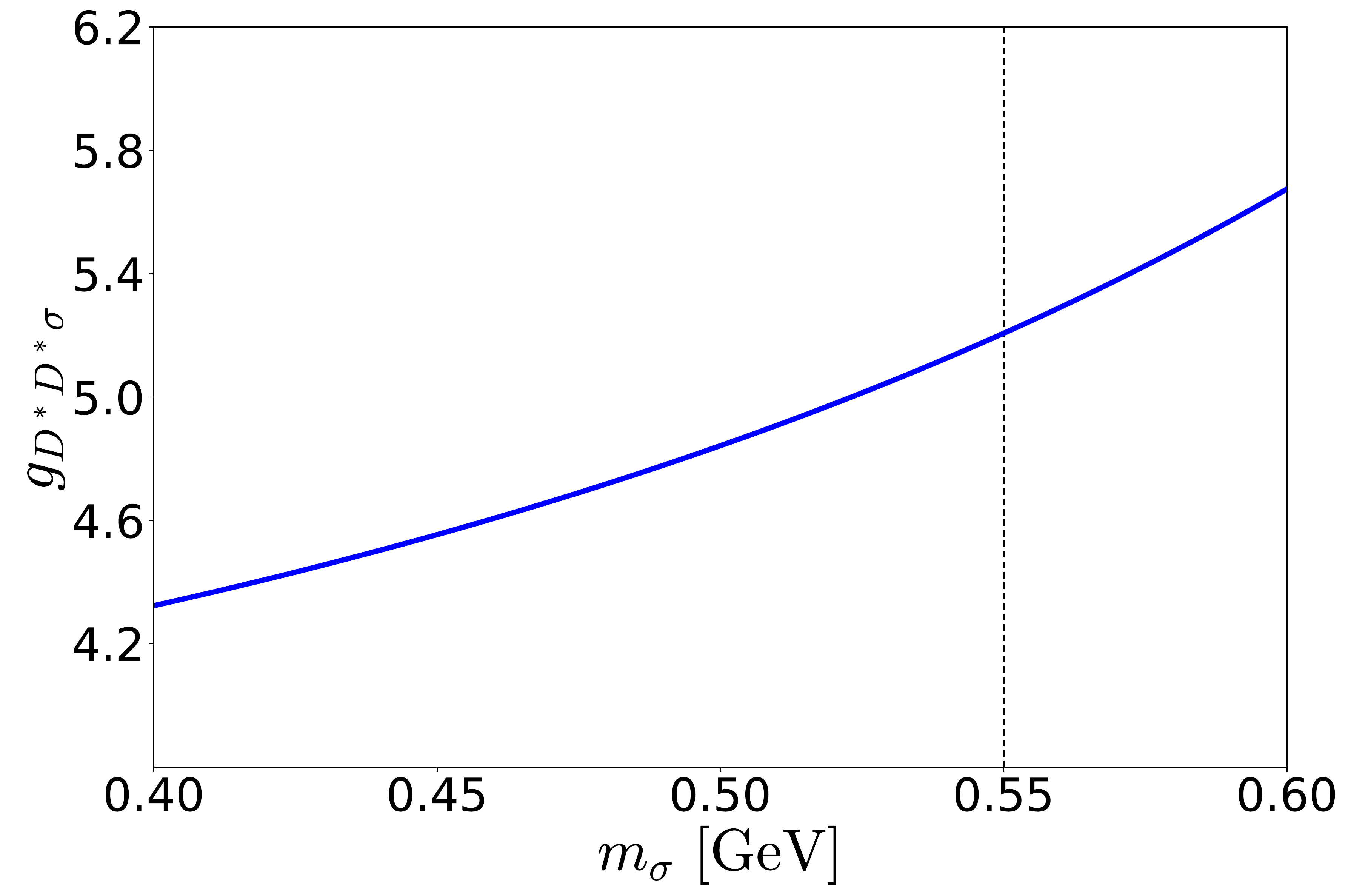}
  \caption{$DD\sigma$ and $D^*D^*\sigma$ coupling constants as a
    function of the $\sigma$-meson mass.}
  \label{fig:9}
\end{figure}
Since the $\sigma$-meson has a broad mass distribution, it is rather
difficult to determine its precise mass. Thus, it is of
great importance to see if the $\sigma$ coupling constants are
sensitive to the value of $m_\sigma$. The left and right panels of
Fig.~\ref{fig:9} draw the dependence of $g_{DD\sigma}$ and
$g_{D^*D^*\sigma}$ on the $\sigma$-meson mass, respectively. The
dashed vertical line represents a preferable value for the
$\sigma$-meson mass, i.e. $m_\sigma=0.55$ GeV, which was often taken
in the $NN$ interactions. The values of $g_{DD\sigma}$ and
$g_{D^*D^*\sigma}$ increase mildly as $m_\sigma$ increases. We will
take the $\sigma$ coupling constants at $m_\sigma=550$ MeV as our
final results. 

\begin{figure}[htp]
\centering
\includegraphics[scale=0.5]{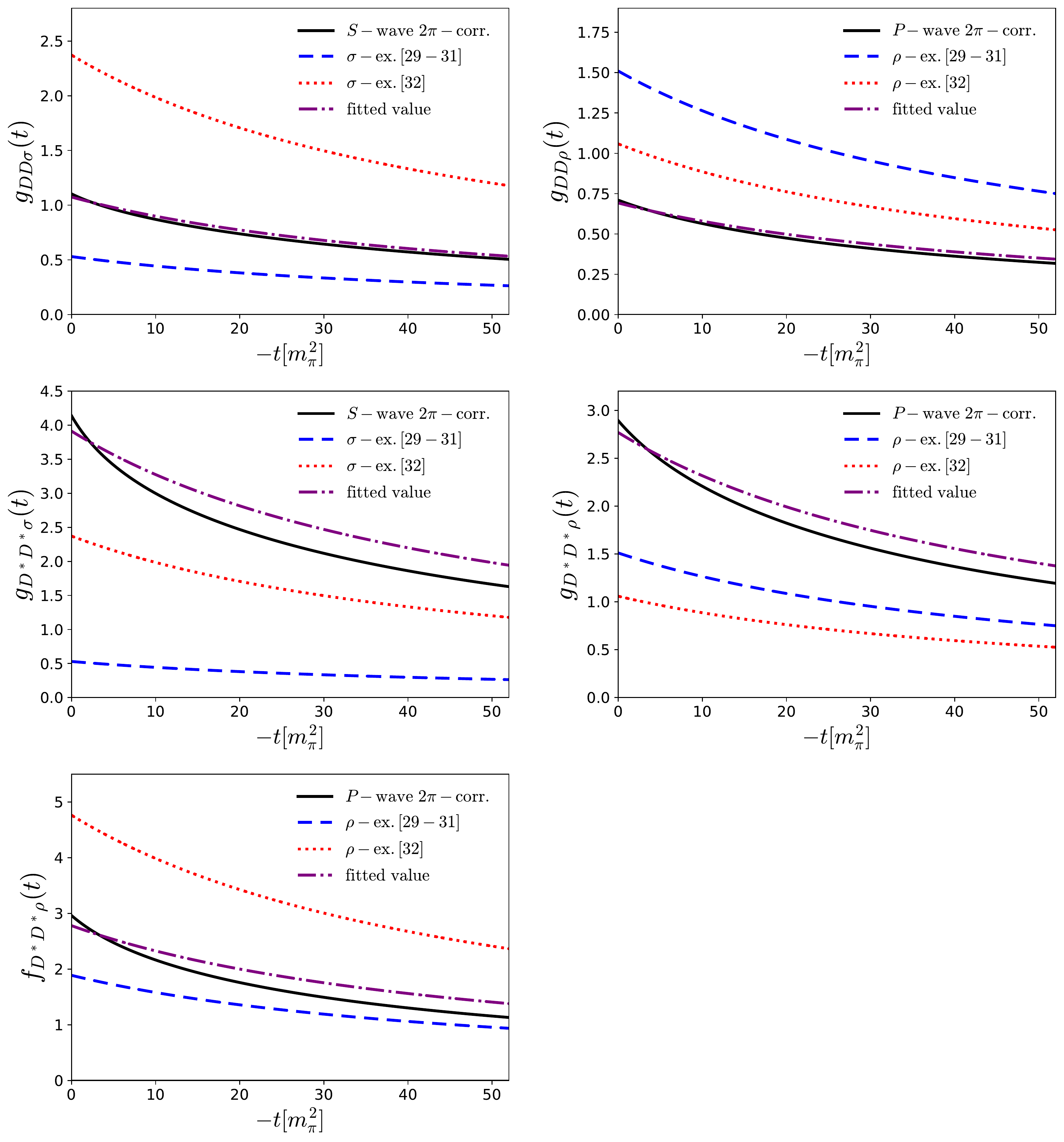}
  \caption{Numerical results for the $\sigma$ and $\rho$ vertex
    functions with correlated $2\pi$ exchange in comparison with
    those with monopole-type form factors. The solid curves draw the
    present results defined in Eq.~\eqref{eq:corre2pi_ff} whereas
    dashed and dotted ones show the $t$ dependence of the vertex
    functions produced by using the coupling constants given in 
  Refs.~\cite{Liu:2008tn, Lee:2009hy, Liu:2010xh} and
  Ref.~\cite{Liu:2019stu}, respectively. The dot-dashed curves
  illustrate the fitted coupling constant by using the monopole-type form
    factor. 
  } 
  \label{fig:10}
\end{figure}
As mentioned previously, the values of the coupling constants from
lattice QCD and other works are often derived at $t=0$, which are off
mass-shell. In the present work, we can only obtain the
\emph{on-mass-shell} coupling constants. In order to extend the
present results to the off-mass-shell region, $t\le 0$, we need to
utilize phenomenologically monopole-type form factors, which are
often employed by various works in studying properties of the exotic
heavy mesons in  meson-exchange pictures. Thus, we introduce a form 
factor given in Eq.~\eqref{eq:corre2pi_ff}, 
which is reduced to the usual monopole-type form factor with the pole
approximation. Though Eq.~\eqref{eq:corre2pi_ff} is a phenomenological
one, it is still very useful for the comparison of the present results
with those from other works at $t=0$. In Fig.~\ref{fig:10}, we depict 
the numerical results for the vertex functions of the $\sigma$ and
$\rho$ mesons both in the $DD$ and $D^*D^*$ channels as functions of
the squared momentum transfer $-t$, comparing them obtained by using
the $\sigma$ and $\rho$ coupling constants taken from 
Refs.~\cite{Liu:2008tn,Lee:2009hy,  Liu:2010xh, Liu:2019stu}.
In the first panel of Fig.~\ref{fig:10}, we draw the present result for
$g_{DD\sigma}(t)$ in the solid curve, while the dashed and dotted ones
correspond to Refs.~\cite{Liu:2008tn, Lee:2009hy, Liu:2010xh} and
Ref.~\cite{Liu:2019stu}, respectively. The dot-dashed curves
illustrate the vertex functions with the present values of $\sigma$
and $\rho$ coupling constants.
Note that the \emph{off-mass-shell} coupling constants are
reduced by approximately $30\,\%$. This can be easily understood by
the following value of the monopole-type form factor
\begin{align}
  \label{eq:2}
\frac{\Lambda_{\sigma}^2-m_\sigma^2}{\Lambda_\sigma^2}  \approx 0.7.
\end{align}
We have similar conclusions also on the $DD\rho$ and $D^*D^*\rho$
coupling constants. However, since the mass of the $\rho$ meson is
about $770$ MeV, the $\rho$ coupling constants are reduced by about
$60\,\%$. Thus, we summarize the present results for the
\emph{off-mass-shell} $\sigma$ and $\rho$ coupling constants at $t=0$
as follows:  
\begin{align}
  \label{eq:sigma_at_0}
g_{DD\sigma}(0)&= 1.07,\;\;\; g_{D^*D^*\sigma}(0)=3.91,  \\
  \label{eq:rho_at_0}
  g_{DD\rho}(0) &=0.69,\;\;\; g_{D^*D^*\rho} (0) = 2.77,\;\;\;
  f_{D^*D^*\rho}(0) = 2.78.
\end{align}
The results for the $g_{DD\rho}$ and
$g_{D^*D^*\rho}$ from lattice QCD~\cite{Can:2012tx}  at $t=0$ are
given as   
\begin{align}
g_{DD\rho}(0) = 4.84(34),\;\;\; g_{D^*D^*\rho}(0) =
  5.94(56),\;\;\;\mbox{lattice QCD~\cite{Can:2012tx}} . 
\end{align}
Comparing the present values with those from other works, we find that
the present results are underestimated. The results from
Refs.~\cite{Bracco:2001dj, El-Bennich:2016bno} are larger than the
present value for the $DD\rho$ coupling constant:
\begin{align}
 g_{DD\rho}(0) &=2.9,\;\;\; \mbox{QCD sum rule~\cite{Bracco:2001dj}}
                 ,\cr 
 g_{DD\rho}(0) &= 6.37,\;\;\; \mbox{Dyson-Schwinger
  approach~\cite{El-Bennich:2016bno}} .     
\end{align}

\begin{table}[htp]
\centering
\caption{$\sigma$ and $\rho$ coupling constant for the $B$ and $B^*$ 
  mesons. In the second column, the present results are listed.  The
  third and fourth ones list the results from Refs.~\cite{Liu:2008tn,
    Lee:2009hy, Liu:2010xh} and Ref.~\cite{Liu:2019stu},
  respectively. 
}
\label{tab:2}
\begin{tabular}[t]{c|ccc}
\hline\hline
& Present work & \cite{Liu:2008tn,Lee:2009hy, Liu:2010xh}
  & \cite{Liu:2019stu}\\  
\hline
$g_{BB\sigma}$ & $7.05$ & $0.76$ & $3.4$ \\
$g_{BB\rho}$ & $8.92$ & $3.71$ & $2.6$\\
$g_{B^*B^*\sigma}$ & $9.47$ & $0.76$ & $3.4$ \\
$g_{B^*B^*\rho}$ & $10.1$ & $3.71$ & $2.6$ \\
$f_{B^*B^*\rho}$ & $29.8$ & $4.64$ & $11.7$ \\
\hline\hline
\end{tabular}
\end{table}
It is straightforward to compute the $\sigma$ and $\rho$ coupling
constants for the $B$ and $B^*$ mesons because of the heavy quark
flavor symmetry. However, there are still subtle points that
arise from the mass difference between the charm and beauty quarks.  
Since the $B$ and $B^*$ mesons are much heavier than the $D$ and $D^*$
mesons, we need to introduce larger values of the cutoff masses,
i.e. $\Lambda=6.1$ GeV, which is about 800 MeV larger than the $B$
meson mass. The amplitudes remain almost stable when the cut-off mass
is changed. Another input for the results given in Table~\ref{tab:2}
is the $BB^*\pi$ coupling constant, of which the value is  
$g_{BB^*\pi}=\frac{2 g}{f_\pi} \sqrt{M_B M_{B^*}}=45$. Compared with
that of $g_{DD^*\pi}=17.9$, $g_{BB^*\pi}$ is approximately 2.5 times
larger. Due to the heavy quark flavor symmetry, $g_{B^*B^*\pi}$ has
the same expression as that of $g_{D^*D^*\pi}$.  

In Table~\ref{tab:2} we list the results on the $\sigma$ and $\rho$
coupling constants for the $B$ and $B^*$ mesons.
We observe that magnitudes of the coupling constants for the
$B$ and $B^*$ mesons are much larger than those of $D$ and $D^*$
mesons in Table~\ref{tab:1}. This can be easily understood from the
fact that the value of the $BB^*\pi$ coupling constant is much larger
than that of $DD^*\pi$. Thus, it is very unlikely that the $\sigma$ 
and $\rho$ coupling constants for the $B$ and $B^*$ mesons are set to
be the same as those for the $D$ and $D^*$ based on the heavy quark
symmetry. In fact, Refs.~\cite{Liu:2008tn, Lee:2009hy, Liu:2010xh, Liu:2019stu} put
them equal to those for the $D$ and $D^*$ mesons. However, the 
present results show that these coupling constants are much
larger than those for the $D$ and $D^*$ mesons as shown in
Table~\ref{tab:2}. Even though theoretical uncertainties of the
present work are considered, we can draw a clear conclusion that the
$\sigma$ and $\rho$ coupling constants for $B$ and $B^*$ mesons should
be taken to be at least larger than those for the $D$ and $D^*$
mesons. It indicates that these large values of the
coupling constants may come into significant role in describing the
molecular states consisting of $B\bar{B}$ and $B\bar{B}^*$.  

\section{Summary and conclusion}
In the present work, we derived the five coupling constants,
i.e., the $DD\sigma$, $DD\rho$, $D^*D^*\sigma$, vector and tensor
$D^*D^*\rho$ coupling constants, having constructed the spectral
functions in both the scalar-isoscalar ($\sigma$) and 
vector-isovector ($\rho$) channels. Starting from the effective
Lagrangians, we first computed the off-shell
$D\bar{D}\to \pi\pi$ and $D^*\bar{D}^*\to \pi\pi$ amplitudes in the
pseudophysical region that is defined in the range of $4m_\pi^2 \le
t \le 52 m_\pi^2$. Then we combined these off-shell Born amplitudes
with the off-shell $\pi\pi$ amplitudes that was evaluated within the
framework of the meson-exchange model known as the J{\"u}lich $\pi\pi$
model, making use of the Blankenbecler-Sugar rescattering
equation. Imposing the two-body unitarity, we were able to evaluate
the spectral functions of correlated $2\pi$ exchange for the
$D\bar{D}$ and $D^*\bar{D}^*$ interactions. The $DD$ and $D^*D^*$
amplitudes with correlated $2\pi$ exchange were derived
by the dispersion relations. We presented the $DD$ and $D^*D^*$
amplitudes with $S$- and $P$-wave correlated $2\pi$ exchange,
comparing them with the corresponding ones with $\sigma$ and $\rho$
meson exchanges, respectively. By reproducing the $DD$ and $D^*D^*$
amplitudes with $S$- and $P$-wave correlated $2\pi$ exchange, we 
obtained the numerical results on the \emph{on-mass-shell} coupling
constants and compared them with those of other works.
Having introduced a monopole-like form factor, we discussed the
coupling constants in the region $t\le 0$. Compared with the lattice
data, the present results are quite smaller than them.
We also examined the dependence of the $\sigma$ couplings on the
$\sigma$-meson mass. We computed the $\sigma$ and $\rho$
coupling constants for the $B$ and $B^*$ mesons for completeness. We
found that these coupling constants for the beauty mesons are much
larger than those for the $D$ and $D^*$ mesons. The reason comes from
the fact that the $BB^*\pi$ coupling constants are much greater than
$DD^*\pi$ ones. It leads to the conclusion that it is unlikely for
the $\sigma$ and $\rho$ coupling constants for the $B$ and $B^*$
mesons to be the same as those for the charmed mesons. 

The present results can be used in any one-boson exchange model for
the description of the exotic heavy mesons as weakly bound molecular
states such as the $X(3872)$ exotic mesons. Though the two-boson
exchange may be considered to be small, the effects of the correlated
$2\pi$ exchange may play a very important role in understanding those
exotic mesons. In particular, considering the fact that the
$\sigma$-meson exchange is the main source for the attraction between
heavy mesons such as $D$ and $D^*$ as in the case of the $NN$
interactions, the present determination of the $\sigma$ couplings for
heavy mesons will be rather useful for understanding the exotic heavy
mesons. We can also determine the $\sigma$ and  $\rho$ couplings for
the other processes such as the $D_1 \bar{D}^*$ and $D_s \bar{D}^*$
interactions. The corresponding works are under way.  

\section*{Acknowledgments}
The present work is supported by Inha University Research Grant in
2019 (No. 61489-01).

\begin{appendix}
\section{The derivation of the spetral functions for the $D^*D^*$
  channel}  \label{app:a}
The projection operators for the vector and tensor coupling constants  
in the $s$ channel are defined as
\begin{align} \label{}
\mathcal{A}^{\mu \nu \alpha \beta} &= \frac1{4} g^{\mu \alpha} g^{\nu \beta}, \cr
\mathcal{B}^{\mu \nu \alpha \beta} &= \frac1{2\sqrt{3}t} (g^{\alpha \nu} k^\mu k^\beta
- g^{\alpha \beta} k^\mu k^\nu - g^{\mu \nu} k^\alpha k^\beta
+ g^{\beta \mu} k^\alpha k^\nu), \cr
\mathcal{C}^{\mu \nu \alpha \beta} &= \frac1{4\sqrt{t(4M_{D^*}^2-t)}}
(-g^{\nu \beta} k^\mu P_2^\alpha + g^{\nu \beta} P_2^\mu k^\alpha
- g^{\mu \alpha} P_1^\nu k^\beta + g^{\mu \alpha} k^\nu P_1^\beta),
\end{align}
where
\begin{align} \label{}
k = p_1 - p_3 = p_4 - p_2,\;\;\; P_1 = p_1 + p_3,\;\;\;
P_2 = p_2 + p_4.
\end{align}
In the $t$ channel, they can be reexpressed as
\begin{align} \label{}
\bar{\mathcal{A}}^{\mu \nu \alpha \beta} &= \frac1{4} g^{\mu \alpha}
                                           g^{\nu \beta}, \cr
\bar{\mathcal{B}}^{\mu \nu \alpha \beta} &= \frac1{2\sqrt{3}t}
                                           (g^{\alpha \nu} P^\mu P^\beta
- g^{\alpha \beta} P^\mu P^\nu - g^{\mu \nu} P^\alpha P^\beta
+ g^{\beta \mu} P^\alpha P^\nu), \cr
\bar{\mathcal{C}}^{\mu \nu \alpha \beta} &= \frac1{4\sqrt{t(4M_{D^*}^2-t)}}
(-g^{\nu \beta} P^\mu \bar{P}_2^\alpha + g^{\nu \beta} \bar{P}_2^\mu P^\alpha
- g^{\mu \alpha} \bar{P}_1^\nu P^\beta + g^{\mu \alpha} P^\nu \bar{P}_1^\beta),
\end{align}
where
\begin{align} \label{}
P = p_1 + \bar{p}_3 = \bar{p}_2 + p_4, \;\;\;\bar{P}_1 = p_1 -
    \bar{p}_3, \;\;\; \bar{P}_2 = p_4 - \bar{p}_2.
\end{align}
The imaginary parts of the $D^*\bar{D}^*$ amplitudes are written as
\begin{align} \label{}
\mathrm{Im}\,\mathcal{M}_{D^*D^*\sigma} &=
16 M_{D^*}^2 \epsilon_\mu(\mathbf{p},\lambda_1)
  \epsilon_\alpha(-\mathbf{p},\lambda_3)
\epsilon_{\nu}^*(-\mathbf{p'},\lambda_2)
 \epsilon_{\beta}^*(\mathbf{p}',\lambda_4) \bar{\mathcal{A}}^{\mu \nu
 \alpha \beta}
 \rho_{00}^{(+),1}(t), \cr
 &= \mathrm{Im}\, p_1^{+,J=0} (t)
\end{align}
and
\begin{align} \label{}
\mathrm{Im}\,\mathcal{M}_{D^*\bar{D}^*\rho} &=
\epsilon_\mu(\mathbf{p},\lambda_1) \epsilon_\alpha(-\mathbf{p},
  \lambda_3) \left[\rho_{11}^{(-),1}(t) 4(s-u)
\bar{\mathcal{A}}^{\mu \nu \alpha \beta} \right.\cr
&\left.+\, \rho_{11}^{(-),2}(t) 32\sqrt{3}t \,
\bar{\mathcal{B}}^{\mu \nu \alpha \beta}
+ \rho_{11}^{(-),3}(t) 16\sqrt{t(4M_{D^*}^2-t)} \,
\bar{\mathcal{C}}^{\mu \nu \alpha \beta}
\right]\epsilon_{\nu}^*(-\mathbf{p'},\lambda_2)
 \epsilon_{\beta}^*(\mathbf{p}',\lambda_4), \cr
&= 2\,\mathrm{Im} p_1^{-,J=1}(t) d_{00}^1(\cos{\theta})
+ 2\,\mathrm{Im} p_2^{-,J=1}(t) d_{1,-1}^1(\cos{\theta})
+ 2\,\mathrm{Im} p_3^{-,J=1}(t) d_{0,1}^1(\cos{\theta}).
\end{align}
The $p_i^{\pm, J}$ are defined as
\begin{align} \label{}
\mathrm{Im}\,p_1^{+,J=0} &\equiv \frac1{64\pi} \sqrt{\frac{t-4m_\pi^2}{t}}
|\mathcal{M}_{D^*\bar{D}^*\to\pi\pi}^{(+),J=0}(1,1)|^2, \cr
\mathrm{Im}\,p_1^{-,J=1} &\equiv \frac1{64\pi} \sqrt{\frac{t-4m_\pi^2}{t}}
|\mathcal{M}_{D^*\bar{D}^*\to\pi\pi}^{(-),J=1}(1,1)|^2, \cr
\mathrm{Im}\,p_2^{-,J=1} &\equiv \frac1{64\pi} \sqrt{\frac{t-4m_\pi^2}{t}}
|\mathcal{M}_{D^*\bar{D}^*\to\pi\pi}^{(-),J=1}(1,0)|^2, \cr
\mathrm{Im}\,p_3^{-,J=1} &\equiv \frac1{64\pi} \sqrt{\frac{t-4m_\pi^2}{t}}
\mathrm{Re}\left\{\mathcal{M}_{D^*\bar{D}^*\to\pi\pi}^{(-),J=1 \dagger}(0,1)
\mathcal{M}_{D^*\bar{D}^*\to\pi\pi}^{(-),J=1}(1,1) \right\}.
\end{align}
After subtraction of the Born amplitudes, we find the spectral
functions as follows:
\begin{align}
\rho_{00}^{(+),1}(t) &= \frac{3}{4M_{D^*}^2}\left[\mathrm{Im}\,p_1^{+,J=0}(t) -
  \mathrm{Im}\,p_{1,\mathrm{Born}}^{+,J=0}(t)\right],\cr
\rho_{11}^{(-),1}(t) &= \frac{2}{4M_{D^*}^2-t}\left[\mathrm{Im}\,p_1^{-,J=1}(t) -
  \mathrm{Im}\,p_{1,\mathrm{Born}}^{-,J=1}(t)\right],\cr
\rho_{11}^{(-),2}(t) &= \frac{2M_{D^*}^2}{t(4M_{D^*}^2-t)}
                       \left[\mathrm{Im}\,p_2^{-,J=1}(t) -
  \mathrm{Im}\,p_{2,\mathrm{Born}}^{-,J=1}(t)\right],\cr
\rho_{11}^{(-),3}(t) &= \frac{2M_{D^*}}{4\sqrt{t}(4M_{D^*}^2-t)}
                       \left[\mathrm{Im}\,p_3^{-,J=1}(t) -
  \mathrm{Im}\,p_{3,\mathrm{Born}}^{-,J=1}(t)\right].
\end{align}

\end{appendix}

\end{document}